\documentclass[12pt]{JHEP3}
\usepackage{epsfig}
\usepackage{amsmath}
\newcommand{\p}{\partial}
\newcommand{\nn}{\nonumber}

\newcommand{\tb}{\overline{T}}
\title{
\makebox[150mm]{Boundary String Field Theory as a Field Theory}
--- Mass Spectrum and Interaction
}
\author{
Koji Hashimoto$^*$ and Seiji Terashima$^\dagger$\\ 
$^*$Institute of Physics, University of Tokyo\\
Komaba, Tokyo 153-8902, Japan\\
E-mail: \email{koji@hep1.c.u-tokyo.ac.jp}\\
$^\dagger$New High Energy Theory Center, Rutgers University\\
126 Frelinghuysen Road, Piscataway, NJ 08854-8019, USA\\
E-mail: \email{seijit@physics.rutgers.edu}\\
}

\abstract{
We study the BSFT actions by using an analytic continuation in momentum
space. We compute various two- and three- point functions for some
low-lying excitations including {\it massive states} on BPS/non-BPS
D-branes. The off-shell two-point functions for the tachyon, the gauge 
field and the massive fields are found to reproduce the well-known
string mass-shell conditions. We compare our action with the tachyon
actions previously obtained by the derivative expansion (or the linear
tachyon profiles), and find complete agreement. Furthermore, we
reproduce the correct on-shell value of the tachyon-tachyon-gauge
three-point function on brane-anti-brane systems. Though inclusion of
the massive modes has been thought difficult because of the
non-renormalizability in string $\sigma$ models, we overcome this by
adopting general off-shell momenta and the analytic continuation.
}
 

\preprint{
{\normalsize{\tt hep-th/0408094}}\\ 
{\normalsize UT-Komaba/04-9}
}

\begin{document}

\section{Introduction}

The outstanding problems in string theory, such as determination of its
vacuum, require complete non-perturbative definition of string/M theory. 
At present, only a few candidates for it are in our hand, among which
the most promising formulation may be string field theories
\cite{BSFT,Wittencubic, HIKKO+Berkovits}. However, their techniques
developed for calculating various physical quantities are yet not
enough for checking the validity and efficiency of the string field
theories. Background-independent (or boundary) string field theory
(BSFT) \cite{BSFT} is one of the formulations of the string field
theories. For superstrings, the BSFT action $S$ is conjectured to be
simply given by the partition function $Z$ of supersymmetric boundary
$\sigma$ models \cite{KMM2,ZS,Andreev-Tseytlin}. The BSFT turned out to
be very useful in describing exactly some of the important consequences
of the tachyon condensation based on the Sen's conjectures
\cite{Senconje} such as tensions of topological defects. Very
few have been as successful as the BSFT in extraction of off-shell
information in string theory, which suggests that the BSFT is an
essential touchstone for exploring non-perturbative aspects of string
theories and even obtaining a proper definition of string/M theory.  

Though the BSFT is a very attractive formulation, it has some
shortcomings to be overcome. One of those is that the definition of the
BSFT action given in \cite{BSFT} is rather formal and it is difficult to
apply it for actual purposes. Consequently, its relation to worldsheet
spectra and string scattering amplitudes is rather unclear. This
relation is one of the main points we are going to clarify in this
paper. One of the other problems is difficulty in incorporating string 
massive modes in the BSFT. Since the BSFT action is defined with
boundary perturbations (interactions) of the worldsheet $\sigma$ model
action, there is apparent difficulty in defining its partition function
for non-renormalizable boundary perturbations. More explicitly, since
$\frac{\p}{\p\tau}X(\tau)$ has mass dimension one in the boundary one
dimensional theory, only the tachyon $T(X)$ and the massless gauge fields
$A_\mu(X)$ are renormalizable couplings of the boundary $\sigma$ model,
once perturbative expansion in powers of dimensionless $X$ is employed
(this corresponds to Taylor (or derivative) expansion of the space-time
fields in the boundary couplings). For this reason, it has been widely
believed 
that only for the tachyon and the gauge field one can compute BSFT
actions, but one can not compute the action for massive modes since they
correspond to non-renormalizable boundary perturbations, of which the
mass dimensions are greater than one. However, there is a loophole in
this argument: we may include an infinite number of boundary couplings
so that the renormalization is consistent \cite{BM}. Although it seems
difficult 
naively to find such a set of bases of infinite dimensions, our answer
is quite simple --- we use a Fourier decomposition instead of the Taylor
expansion, for the boundary interactions.  For example, if  we express
$T(X(\tau))$ in the normal-ordered plane-wave basis as 
$T(X)=\int\!dk\;  t(k) :\!e^{i k\cdot X(\tau)}\!\!:\;$,  
the dimension of this normalized field is less than one for an
appropriate region of the momentum value $k^2$ (this procedure was
briefly discussed in appendix A of \cite{KMM} and similar methods were
used in \cite{Frolov, Coletti}). As we will see in this paper, if we
consider $t(k)$ with appropriately negative $k^2$, the partition
function $Z$ is computed to be  finite and {\it there is no need to
perform the renormalization further}, thus the BSFT action is well
defined for perturbatively all order in $t(k)$. Moreover, we can
analytically continue the BSFT action to include $t(k)$ with any
$k$. The important point is that {\it this procedure can be applied also
for the massive fields} and we have a BSFT action for all (bosonic)
space-time fields. With use of the normal ordered Fourier basis and the
analytic continuation, the BSFT (and also boundary $\sigma$ models) can
incorporate all the excitations of the open string including the massive
modes with arbitrary momenta. 

In this paper, we study the BSFT action using this analytic continuation
method. We compute various two- and three- point functions for some
low-lying excitations (including massive states) on BPS/non-BPS
D-branes and brane-anti-branes. In this paper we assume $S_{\rm
BSFT}=Z$ \cite{ZS}. The off-shell two-point functions for the tachyon,
the gauge field and the massive fields are found to be reproducing 
the well-known string mass-shell conditions. We compare our action with
the tachyon action previously obtained in \cite{KMM2,KL,TTU} by the
derivative expansion (or linear tachyon profiles), and find complete
agreement for terms existing in both. This is an interesting result
since the peturbative tachyon mass was not correctly obtained previously
in \cite{KMM2,Tseysigma} becasuse of truncation of higher derivative
terms in the boundary interaction, while our analytic continuatinuation
method includes arbitrary higher drivative terms and gives the correct 
perturbative tachyon mass. This mechanism has been applied to the
tachyon condensation senario in our previous paper \cite{pre} in which
we have showed that the BSFT linear tachyon backgrounds corresponding 
to lower-dimensional BPS D-branes and rolling tachyons in fact give
rise to deformations of mass-shell structures (expected in effective
field theory results such as in \cite{GHY}) for all the open string
excitations, which verifies the Sen's conjectures
\cite{Senconje}. One of the other  intriguing aspects of our two-point 
functions is that they seem to require a cut-off for large momenta ---
this might be the appearance of the minimum length in string theory. 

For three-point functions, we reproduce the correct on-shell value of
the tachyon-tachyon-gauge three-point function on brane-anti-brane
systems. (Most of the other combinations of the fields vanish in the
superstring case.) To compute the on-shell value of the three-point
function, the limiting procedure empoloyed in \cite{Frolov, Coletti} is
not correct in general. Furthermore, \cite{Frolov} pointed out a problem
in which the three-point function with a gauge field exhibits a
divergence. We solve these problems by using a field redefintion of the
space-time fields which makes the three-point function regular at the
on-shell value. The three-point function obtained in this way coincides
with the S-matrix computation. As a property of interactions in the
BSFT, the following consistency condition is expected \cite{Frolov}: 
the action $S=Z$ already includes the vertices which reproduce
on-shell tree level S-matrix \cite{Andreev-Tseytlin}, while the path
integral of $e^{iS}$ or one-particle-reducible Feynman graphs would give
further contributions to it, thus these additional contributions should
vanish with on-shell external legs. We have checked this explicitly for
our three-point function.

The organization of the paper is as follows. 
In section \ref{sec-analytic}, after describing our general procedures
of computation of the BSFT action, we give two-point functions of
spacetime fields including massive excitations reproducing correct
superstring spectra for BPS/non-BPS D-branes. A detailed consistency
with the tachyon/gauge BSFT actions in the previous literatures is
studied. In section \ref{sec-3}, a three-point function on a non-BPS
brane is obtained and its properties, especially the reproduction of a
string scattering amplitude, are studied. section \ref{sec-gene} is for
discussing various interesting properties of the obtained off-shell
action, including an indication of a spacetime resolution in string
theory. In appendix \ref{app:rolling} we apply our action to reproduce a
part of Sen's result on rolling tachyons \cite{originalroll, LNT}, and
in appendix \ref{app:constb}, a background gauge field strength is
introduced to see derivative corrections to Born-Infeld actions. In this
paper we consider only superstrings since their BSFT action is simple.


\section{BSFT action via analytic continuation}
\label{sec-analytic}

After pointing out the difficulty in getting mass-shell conditions in
previous techniques (Taylor expansions of the boundary couplings) 
in the BSFT, in this section we explain how the different choice of the
expansion basis consistently gives a BSFT action incorporating the
mass-shell conditions. Emphasis is on the fact that, even for massive
excitations which are widely believed to be non-renormalizable and so
difficult 
to be treated in boundary $\sigma$ models, we can successfully write
down the BSFT action via the ``analytic continuation method''. We
utilize the momentum dependence of conformal dimensions of off-shell
vertex operators to obtain a finite partition function which is the BSFT
action. Though one has to make the analytic continuation to get into
physically interesting regions of momenta such as the nearly on-shell
region, this method turns out to be consistent with all known actions of
tachyons and gauge fields derived with Taylor (derivative) expansion of
the boundary couplings. 


\subsection{Brief review of super BSFT action}
\label{review}

The disk partition function for a BPS D9-brane is defined by
\begin{eqnarray}
\label{eq5}
Z &=& \int\! DX D\psi \;\exp[-I(X,\psi)] \ .
\end{eqnarray}
In this definition the $\sigma$ model action $I$ is composed of two
terms, $I = I_0+I_{\rm B}$, where $I_0$ is the bulk part\footnote{
The spacetime metric is $\mbox{diag}(-1,1,1,\cdots,1)$ in this paper.} 
\begin{eqnarray}
I_0 &=& \frac{1}{4\pi}\int_{\Sigma}d^2 z 
\left[\frac{2}{\alpha'} \partial_z X^{\mu}
\partial_{\bar{z}}X_{\mu}+\psi^{\mu}\partial_{\bar{z}}\psi_{\mu}
+\tilde{\psi}^{\mu}\partial_{z}\tilde{\psi}_{\mu}\right],
\label{eq1}
\end{eqnarray}
and $I_{\rm B}$ is a boundary interaction which is written 
by a superfield whose argument is restricted to the boundary,
\begin{eqnarray}
 {\bf X}^\mu(\tau,\theta) 
= X^\mu(\tau) + \sqrt{2\alpha'}i
\theta \psi^\mu(\tau) \ .
\label{xdecomp}
\end{eqnarray}
In this paper we take a convention $\alpha'=2$. When applied to the
context of the BSFT, $\Sigma$ should not be the upper half plane but
a unit disk, and thus $\p\Sigma$ is not the real axis but a unit
circle. So, $\tau$ parametrizes the boundary circle of the disk, 
$-\pi < \tau \leq \pi$.
As an example of the boundary perturbation, a massless gauge field on a
BPS D-brane is represented by  
\begin{eqnarray}
I_{\rm B} &=& \int_{\partial\Sigma}\!\!\!d\tau d\theta
[-iD_{\theta}{\bf X}^{\mu}A_{\mu}({\bf X})] \ ,
\label{gaugeint}
\end{eqnarray}
where the superspace derivative is 
\begin{eqnarray}
D_{\theta} &=& \frac{\partial}{\partial\theta}
+\theta\frac{\partial}{\partial\tau} \ .
\label{supersd}
\end{eqnarray}
Other space-time fields can be incorporated as boundary couplings
possessing higher powers of $D_\theta$. 

For a brane-anti-brane system, we should include the following boundary
fermion which represents the Chan-Paton factor of the brane-anti-brane
\cite{WHKM}. In the superfield notation, the boundary fermion is 
\begin{eqnarray}
{\bf \Gamma}(\tau,\theta) &=& \eta(\tau)+\theta F(\tau) \ ,
\label{bfer}
\end{eqnarray}
and its kinetic action is given by \cite{KL,TTU}
\begin{eqnarray}
I_{\bf \Gamma} &=& \int_{\partial\Sigma}\!\!d\tau d\theta\;
[-\bar{ {\bf \Gamma}} D_{\theta} {\bf \Gamma}]=
\int_{\partial\Sigma} \!\!d\tau\;
[ \bar{\eta} \dot{\eta} -\bar{F} F] \ .
\end{eqnarray}
This means that $\eta$ is a propagating boundary fermion while $F$ is an
auxiliary field. By the canonical quantization of this action, we may
regard $\bar{\eta}, \eta$ and $[\bar{\eta}, \eta]$ as Pauli matrices
$\sigma_+, \sigma_-$ and $\sigma_3$, respectively, which are Chan-Paton
degrees of freedom of the two branes. The number of ${\bf \Gamma}$'s
in the interaction terms is related to the GSO parity. To obtain The
disk partition function, we should path-integrate also ${\bf \Gamma}$, 
\begin{eqnarray}
Z &=& \int \!DX D\psi D\eta DF \;\exp[-I(X,\psi)
-I_{\bf \Gamma}(\eta,F)] \ . 
\end{eqnarray}

To get a boundary action for a non-BPS D-Brane, we simply restrict 
${\bf \Gamma}$ to a real superfield. As an example, the tachyon field on
the D-brane is represented as 
\begin{eqnarray}
I_{\bf \Gamma} + I_{\rm B} 
&=& \int_{\partial\Sigma}\!\!\!\!\! d\tau d\theta
\left[-{\bf \Gamma} D_{\theta} {\bf \Gamma} + 
\frac{ T({\bf X})}{\sqrt{2 \pi}}  {\bf \Gamma} \right]
\nn\\
&=&
\int_{\partial\Sigma}\!\!\!\!\! d\tau 
\left[ \eta \dot{\eta} -F^2+
i \sqrt{\frac{2}{\pi}}\psi^\mu \eta \p_{\mu} T(X)
-\frac{T(X)}{\sqrt{2 \pi}} F \right] \ .
\nonumber 
\end{eqnarray}
In this boundary action, we can easily integrate out the auxiliary field
$F$ to get 
\begin{eqnarray}
&& I_{\bf \Gamma}+I_{\rm B} = \int_{-\pi}^\pi d\tau 
\left[
\eta \dot{\eta} +
i \sqrt{\frac{2}{\pi}}\psi^\mu \eta \p_{\mu}
T(X)
+\frac{1}{8\pi} T(X)^2 
\right].
\label{int1}
\end{eqnarray}
When the tachyon field is constant, we immediately obtain 
the well-known spacetime tachyon potential of the form $e^{-T^2/4}$.

The super BSFT action is conjectured to be given by the disk partition
function 
\begin{equation}
S_{\rm BSFT}= Z \ .
\label{SZas}
\end{equation}
We normalize $Z$ as $Z={\cal T}\int d^{10} x$ when all space-time fields
vanish, where ${\cal T}$ is the tension of the corresponding brane(s).
(Hereafter, we use this normalization of $Z$.) The validity of the
assumption (\ref{SZas}) has been discussed in \cite{ZS}. 

To evaluate the partition function of the open superstring $\sigma$
model, one needs to specify the explicit form of the boundary
couplings. Historically, Taylor expansions for the boundary coupling 
have been adopted in which for example the tachyon field is expanded as 
\begin{eqnarray}
 T(X) = T(x) + \p_\mu T(x) \hat{X}^\mu + \frac12 \p_\mu \p_\nu T(x) 
\hat{X}^\mu \hat{X}^\nu + \cdots \ ,
\label{simpleex}
\end{eqnarray}
where the worldsheet scalar field $X^\mu$ is decomposed into its zero
mode $x^\mu$ plus the oscillator mode $\hat{X}^\mu$. The multiple scalar
fields should 
be normal-ordered so that the boundary coupling is well-defined. The
partition function can be evaluated using the worldsheet propagators
\begin{eqnarray}
 \left\langle \hat{X}^\mu(\tau) \hat{X}^\nu(0)\right\rangle
= -4 \eta^{\mu\nu}\log \left|
2\sin\frac{\tau}{2}
\right| \ , \quad 
\left\langle \psi^\mu(\tau) \psi^\nu (0)\right\rangle
 = \frac{1}{2\sin\frac{\tau}{2}} \ .
\label{prop1}
\end{eqnarray}
The propagator for $\eta$ (in the non-BPS case) is
\begin{eqnarray}
 \left\langle\eta(\tau) \eta(\tau')\right\rangle
= \frac{1}{2}\epsilon(\tau-\tau') = \frac{1}{2}
\frac{\sin(\tau/2)}{|\sin(\tau/2)|} \ ,
\label{prop2}
\end{eqnarray}
where $\epsilon(\tau_{12})$ is the sign function. Due to the worldsheet
periodicity, the last expression is appropriate, if we define 
$\frac{\sin(\tau/2)}{|\sin(\tau/2)|}=0$ for $\tau = 2n \pi$ 
($n \in {\bf Z}$).

The Taylor expansion (\ref{simpleex}) adopted in most of the literature
has provided various interesting results \cite{Andreev-Tseytlin,
KMM,KMM2,Tseysigma,KL,TTU}, 
especially on the tachyon condensation and the Sen's conjectures. For
the tachyon profiles linear in $X$, the boundary interaction is
path-integrated exactly and gives the tensions of lower-dimensional
D-branes. However, this simple expansion (\ref{simpleex}) is obviously 
incompatible with general mass-shell conditions, because basically the
above expansion is a derivative (or $\alpha'$) expansion. Throwing away
the higher derivatives to compute the partition function is valid only
for nearly massless states, though generic on-shell conditions are
apparently different from the massless condition. In this paper, we
adopt a different expansion to reconcile the on-shell condition with the
computability of the partition function.  


\subsection{Tachyon two-point function}
\label{subsec-two-point tachyon}

The essence described in Appendix A of \cite{KMM} is to use a plane-wave
basis instead of the Taylor (derivative) expansion (\ref{simpleex}) to
extract the structure of the off-shell tachyon state which can be nearly
on-shell. We expand the tachyon field as     
\begin{eqnarray}
 T(X) = \int\! dk\; t(k) e^{ik\cdot { X}} \;
= \int\! dk\; t_{\rm R}(k) :e^{ik\cdot { X}}: \;
= \int\! dk\; t_{\rm R}(k)\; e^{ik\cdot x} :e^{ik\cdot \hat{{ X}}}:
\ .
\label{tacdef}
\end{eqnarray}
The function $t(k)$ is a momentum representation of the tachyon field.
We allow generic function $t(k)$ which is thus off-shell. 
We introduced the normal ordering so that the tachyon off-shell coupling
itself is well-defined on the boundary of the worldsheet. The relation
to the renormalized tachyon field is 
\begin{eqnarray}
 t(k) = t_{\rm R}(k) \exp\left[2k^2 \log\epsilon\right]
\end{eqnarray}
where $\epsilon$ is the regularization parameter in a regularized
version of the propagator (\ref{prop1}), 
\begin{eqnarray}
\left\langle \hat{X}^\mu(\tau) \hat{X}^\nu(0)\right\rangle
=2 \eta^{\mu \nu} \sum_{m\neq 0} 
\frac{1}{|m|} e^{im\tau - |m|\epsilon}
\ . 
\label{regpro}
\end{eqnarray}
Accordingly, the renormalized tachyon field in the coordinate
representation is 
\begin{eqnarray}
T(x)= \exp\left[-2(\log\epsilon)\p_\mu \p^\mu \right]T_{\rm R}(x) \ . 
\label{regt}
\end{eqnarray}

In superstring theories, the open string tachyon appears in unstable  
D-branes. Here we consider the tachyon in a non-BPS D-branes. The
relevant boundary action $I_{\bf \Gamma}+I_{\rm B}$ is
(\ref{int1}). Since we work in component fields in the rest of this
paper, we refer to (\ref{int1}) without the $\eta$ kinetic term,
i.e.~the second and the third term in (\ref{int1}), as $I_{\rm B}$.  
The tachyon $n$-point function is given by
\begin{eqnarray}
\frac{1}{n!} \int\! dx\; \left\langle (-I_{\rm B})^n
\right\rangle \ ,
\label{partt}
\end{eqnarray}
where $\langle \; \rangle$ denotes the path-integration over $\hat{X}$, 
$\psi$ and $\eta$ with the worldsheet action 
$I_0 + \int\!d\tau \;\eta\dot{\eta}$ for the non-BPS case. 
A simplification occurs as explained in \cite{F-Tseytlin}, the term
$T^2$ in the tachyon coupling $I_{\rm B}$ vanishes in an appropriate
region of the momenta, since  
\begin{eqnarray}
&& T(X(\tau))T(X(\tau)) =
\int\!\! dk dk' \; t_{\rm R}(k)t_{\rm R}(k')
e^{i(k+k')\cdot x}
\lim_{\epsilon \rightarrow 0}
:e^{ik\cdot \hat{X}(\tau+\epsilon)}\!:\; 
:e^{ik'\cdot \hat{X}(\tau)}\!:
\nn\\
&&
= 
\int\!\! dk dk' \; t_{\rm R}(k)t_{\rm R}(k')
e^{i(k+k')\cdot x}
\lim_{\epsilon \rightarrow 0}
\exp\left[ 4k\cdot k'\log|\epsilon|
\right] :e^{i(k+k')\cdot \hat{X}(\tau)}\!:
\nn\\
&&
= \int\!\! dk dk' \; t_{\rm R}(k)t_{\rm R}(k')
e^{i(k+k')\cdot x}
\lim_{\epsilon \rightarrow 0}
\epsilon^{4k\cdot k'}:e^{i(k+k')\cdot X(\tau)}\!: \;
\nn\\
&&
=\;0 \ .
\label{vanisht2}
\end{eqnarray}
Here we have assumed $k\cdot k' >0$. Thus we may consider just 
the term $\psi^\mu \eta\p_\mu T$ in $I_{\rm B}$ for the computation of
the relevant part of the partition function (\ref{partt})
and then analytically continue the result to all other region of $k$.
What we are doing here is a kind of an off-shell version of \cite{LNT}.

In this subsection we obtain the tachyon two-point function 
$\int dx  \langle I_{\rm B} I_{\rm B}\rangle$
which provides the mass-shell condition,
with use of the boundary propagators 
(\ref{prop1}), (\ref{prop2}).\footnote{After completion of this work, we
found that this subsection has an overlapping result with
\cite{nishimura}.}
Noting the well-known result 
\begin{eqnarray}
\left\langle 
:e^{ikX}(\tau): :e^{i\tilde{k}X}(0):
\right\rangle
= 
\left|
2\sin \frac{\tau}{2}
\right|^{4k\cdot\tilde{k}} e^{i(k+\tilde{k})x} \ ,
\label{well-known}
\end{eqnarray}
we obtain $\langle I_{\rm B} I_{\rm B} \rangle$ as 
\begin{eqnarray}
\lefteqn{
\int\!\! d\tau_1 d \tau_2
\left\langle 
\left(i \sqrt{\frac{2}{\pi}}\right)
:\psi^\mu\eta \p_{\mu}T(\tau_1):
\left(i \sqrt{\frac{2}{\pi}}\right)
:\psi^\nu \eta \p_{\nu}T(\tau_2):
\right\rangle
}
\nn\\
&&
= 
\int\!\! dk_1dk_2 \int\!\! d\tau_1 d \tau_2
\left(\frac{-2}{\pi}\right) \frac{1}{2}\epsilon(\tau_{12})
\frac{-1}{2\sin(\tau_{12}/2)}
\left|
2\sin\frac{\tau_{12}}{2}
\right|^{4k_1\cdot k_2} \eta^{\mu\nu}
\nn\\
&& \quad \times
e^{ik_1 \cdot x + ik_2 \cdot x}
(ik_1)_\mu t_{\rm R}(k_1) (ik_2)_\nu t_{\rm R}(k_2)
\nn\\
&&
=\int\!\! dk_1 dk_2\; 2^{4k_1\cdot k_2}
\int\!\! d\tau_{12} 
\left|
\sin\frac{\tau_{12}}{2}
\right|^{4k_1\cdot k_2-1} \eta^{\mu\nu}
 e^{ik_1 \cdot x + ik_2 \cdot x}
(-k_1\cdot k_2) t_{\rm R}(k_1)t_{\rm R}(k_2)
\nn\\
&&
= 
\int\!\!dk_1dk_2\; 
e^{ik_1 \cdot x + ik_2 \cdot x}
(-k_1\cdot k_2) t_{\rm R}(k_1)t_{\rm R}(k_2)
2^{4k_1 \cdot k_2} 2\sqrt{\pi} \frac{\Gamma(2k_1 \cdot k_2)}
{\Gamma(2k_1 \cdot k_2 + 1/2)}
\nn\\
&& 
=-
\int\!\!dk_1dk_2\;  e^{i (k_1 + k_2) \cdot x}
 t_{\rm R}(k_1)t_{\rm R}(k_2)
2^{4k_1 \cdot k_2} \sqrt{\pi} \frac{\Gamma(2k_1 \cdot k_2+1 )}
{\Gamma(2k_1 \cdot k_2 + 1/2)} \ .
\end{eqnarray}
Hence defining 
\begin{eqnarray}
Z^{(2)}(y) 
\equiv 
2^{2y}\sqrt{\pi}
\frac{\Gamma(y+1)}
{\Gamma(y+1/2)} \ ,
\label{Z2tac2}
\end{eqnarray}
we obtain the BSFT action 
\begin{eqnarray}
S =Z & = & 
{\cal T}\int\!\! dx
\left[
1 -\frac12 
 \int\!\! dk_1 dk_2\; 
t_{\rm R}(k_1)e^{ik_1\cdot x}
Z^{(2)}(\alpha'k_1\cdot k_2)
t_{\rm R}(k_2)e^{ik_2\cdot x}
+ {\cal O}(t_{\rm R}^4)
\right]
\nn\\
& = & 
{\cal T}\int\!\! dx  
\left[
1 -\frac12
T_{\rm R}
(x)Z^{(2)}(-\alpha'\overleftarrow{\p}\cdot\overrightarrow{\p})
T_{\rm R}(x)
+ {\cal O}(T_{\rm R}^4)
\right] \ ,
\label{tacbsft}
\end{eqnarray}
where we recovered the explicit $\alpha'$ dependence by $\alpha'/2=1$.
Note that here we haven't performed the target space zero-mode integral
$\int\! dx$, and so we actually compute the BSFT Lagrangian density $L$. 
The action has information of all order in its spacetime 
derivatives but for small
magnitude of the tachyon field. We shall see interesting properties of
this action below.

First, the tachyon mass-shell condition should be read from the kinetic
function $Z^{(2)}$. The equation of motion for the tachyon follows from
the action as 
\begin{eqnarray}
Z^{(2)}(\p^2) T_{\rm R}(x)= {\cal O}(T_{\rm R}^3) \ . 
\end{eqnarray}
Plane-wave solution with an infinitesimal magnitude is represented as 
$T_{\rm R}(x) = \lambda e^{ik\cdot x}$ ($\lambda \ll 1$), 
for which the equation of motion reduces to 
\begin{eqnarray}
 Z^{(2)}(-\alpha'k^2)=0 \ .
\label{taconshell}
\end{eqnarray}
This can be solved by $k^2 = 1/2\alpha'$, the tachyon mass-shell
condition. 

\begin{figure}[tp]
\begin{center}
\begin{minipage}{13cm}
\begin{center}
\includegraphics[width=8cm]{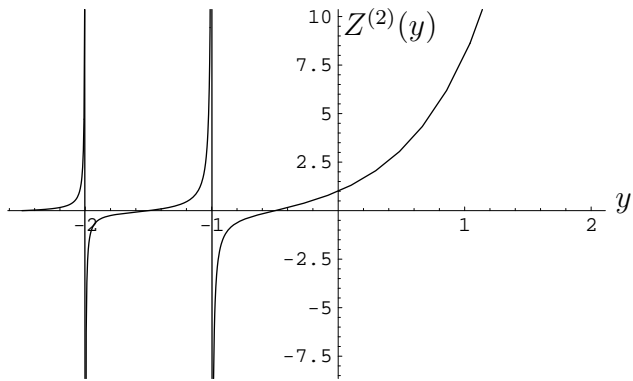}
\put(3,65){$y$}
\put(-100,130){$Z^{(2)}(y)$}
\caption{Behavior of the kinetic function $Z^{(2)}(y)$. It has zeros at 
$y=-n+1/2$ while poles at $y=-n$ ($n\in {\bf Z}^+$). The region $y>-1$
 is regular.}
\label{z2fig}
\end{center}
\end{minipage}
\end{center}
\end{figure}

Strangely, as seen obviously in Fig.~\ref{z2fig}, 
there are an infinite number of solutions for this equation
(\ref{taconshell}),  $\alpha'k^2 = n+1/2$ 
($n \in {\bf Z}_{\geq 0}$). However, poles exists before the next
zero is reached from the regular region $y > 0$. These poles might be
interpreted as a spacetime resolution or a minimum length in string
theory. We shall discuss this point in detail in section
\ref{sec-gene}. 

The kinetic operator $Z^{(2)}(y)$ can be expanded around the on-shell
momentum, in terms of $y+1/2$. The coefficient 
\begin{eqnarray}
\left[
\frac{\p}{\p y} Z^{(2)}(y)
\right]_{y=-\frac12}=\frac{\pi}{2}
\label{coefz2}
\end{eqnarray}
gives us an information on the normalization of the tachyon field once
we require the canonical normalization of the kinetic term. (The terms
higher in the power of $(y+1/2)$ in the expansion of $Z^{(2)}(y)$ can be
absorbed into the field redefinition of the  
tachyon field.) This normalization is necessary in comparing the BSFT
results with known string scattering amplitudes, which will be studied
in section \ref{sec-3}. 


\subsection{Relation to derivative-expanded BSFT tachyon action}

We can compare our result (\ref{tacbsft})
with the known expression for the tachyon
Lagrangians derived so far in BSFT's. An interesting fact is that,
although the definition of the partition function naively suggests that
a constant $T$ gives vanishing result (\ref{vanisht2})
(this is due to the fermion integral $d\theta$ in \cite{LNT}), 
the ``potential'' $T^2$ term is reproduced in our 
result (\ref{tacbsft}) after the analytic continuation. In fact, 
the two-point function in (\ref{tacbsft}) has a regular Taylor expansion 
for small momentum as 
\begin{eqnarray}
&& Z^{(2)}(y)= \sum_{n=0}^\infty a_n y^n 
\ ,  \\
&& a_0 = 1 \ , \quad a_1 = 4\log 2 \ , \quad a_2 = 
-\frac{\pi^2}{6}+8(\log 2)^2 \ , \cdots 
\end{eqnarray}
and in particular, $a_0$ is non-vanishing.
This means that the Lagrangian is expanded in terms of the derivatives 
(i.e.~slowly-varying field approximation) as 
\begin{eqnarray}
\frac{L}{\cal T}= 
1
-\frac{1}{2} T_{\rm R}^2 + 2 \alpha'\log 2 (\partial_\mu T_{\rm R})^2 + 
\left(\!\frac{\pi^2}{12}\!-\!4(\log 2)^2\!\right) {\alpha'}^2
(\p_\mu \p_\nu T_{\rm R})^2
+ {\cal O}(\p^6) \ . 
\label{explag}
\end{eqnarray}
So there appears $T_{\rm R}^2$ ``potential'' term.
Note that, on the other hand, in the conventional BSFT approaches
\cite{KMM2,TTU} the $T_{\rm R}^2$ term was due to the expansion 
of the tachyon potential $e^{-T_{\rm R}^2/4}$ coming from the contact
term $T_{\rm R}(X)^2$ in the boundary action which we have neglected as
seen in (\ref{vanisht2}). Technically speaking, the reason why the
potential term $T_{\rm R}^2$ appears is that we have evaluated the
worldsheet partition function for generic tachyon momentum $k_\mu$ and
taken a limit $k\rightarrow 0$, which is different from the case of
putting $k=0$ from the first place.  We have unawares used an analytic
continuation to evaluate the kinetic operator in regions of interest. 

A conventional string $\sigma$ model approach (Taylor expansion 
or derivative ($\alpha'$) expansion of the boundary couplings) provides
a similar form of the effective action for the tachyon \cite{TTU}, 
\begin{eqnarray}
 \frac{L}{\cal T} 
= 1-\frac12\tilde{T}_{\rm R}^2
+2\alpha'\log 2 (\p_\mu \tilde{T}_{\rm R})^2 
+  {\alpha'}^2 \gamma_0 (\p_\mu\p_\nu \tilde{T}_{\rm R})^2 
+ {\cal O}(\p^6) \ . 
\label{sigmaLt}
\end{eqnarray}
The original result in \cite{TTU} was for a brane-anti-brane, but
restricting the complex tachyon field to be real 
($\frac1{\sqrt{2}}T_{\rm R}^{\rm non-BPS} 
= {\rm Re}\;T_{\rm R}^{\rm D\overline{D}}$), 
the action reduces to
that of the non-BPS brane as above.
The value of the constant $\gamma_0$ defined in \cite{TTU}
can be computed\footnote{
With the following relation for a small $\epsilon$, 
\begin{eqnarray}
&& \sum_{m,r>0} \frac{e^{-(r+m)\epsilon}}{m(m\!+\!r)}
\!= \!\!
\!\! \sum_{m,r>0} \int_\epsilon^\infty\!\!\!\!\!\!\! d\epsilon'
\;\frac{e^{-(r+m)\epsilon}}{m}
= \!
\int_\epsilon^\infty\!\!\!\!\!\!\! d\epsilon'
\frac{-e^{-\frac{\epsilon'}{2}}\log(1\!-\!e^{-\epsilon'})}
{1-e^{-\epsilon'}}
= 
\frac{(\log\epsilon)^2}{2}\! +\! \frac{\pi^2}{6}\! -\! 2 (\log 2)^2 
\!+\! {\cal O}(\epsilon)
\nonumber
\end{eqnarray}
where the summation index 
$m$ is for positive integers while $r$ is for positive half-odd
numbers, $\gamma_0$ can be explicitly evaluated as 
\begin{eqnarray}
 \gamma_0 &=& 
-\lim_{\epsilon \rightarrow 0}
\sum_{m,r>0}\frac{1}{m} \left(\frac{1}{r+m}-\frac{1}{r-m}\right)
e^{-(r+m)\epsilon}
+ \frac{\pi^2}{3} - 4 (\log 2)^2
\nonumber \\
&=&
-2\lim_{\epsilon \rightarrow 0}
\sum_{m,r>0}\frac{1}{(m+r)(m-r)}
e^{-(r+m)\epsilon}
+ \frac{\pi^2}{3} - 4 (\log 2)^2
\;=\;
\frac{\pi^2}{12} - 4 (\log 2)^2 \ .
\end{eqnarray}
}
to be  $\gamma_0 = \pi^2/12 - 4 (\log 2)^2$. This Lagrangian
(\ref{sigmaLt}) completely coincides with our result (\ref{explag}).  
Although it looks that these two Lagrangians were obtained in 
different regularization schemes, in fact these two regularizations turn
out to be the same : the renormalization used in the $\sigma$ model
approach \cite{TTU} was
$T=\tilde{T}_R+\alpha'\log\epsilon \p_\mu \p^\mu \tilde{T}_{\rm R}
+\frac12{\alpha'}^2(\log\epsilon)^2 (\p_\mu \p^\mu)^2\tilde{T}_{\rm R}
+\cdots$ 
with the regularized propagator (\ref{regpro}), 
which coincides with the expansion of our regularization
(\ref{regt}). Therefore we find $T_{\rm R} = \tilde{T}_{\rm R}$ up to
the present order of the derivatives.\footnote{Note that a change of a
renormalization constant, like $\epsilon_R$ in \cite{Frolov},
corresponds to a field redefinition 
$T \rightarrow e^{c \p^2 } T=T+c \p^2 T+\cdots$. 
This gives an extra factor $e^{2 c \p^2 }$ to $Z^{(2)}$,
which does not change the on-shell condition. Thus it is clearly
unphysical.}  

Furthermore, we note that our result (\ref{explag}) is consistent 
also with the ``usual'' BSFT action  
\begin{eqnarray}
S = {\cal T}\!
\int \! dx \;e^{-T^2}\! {\cal F}(2 \alpha' \p^\mu T \p_\mu T) \ ,
\;  
{\cal F}(x)\equiv x \frac{4^x}{2} \frac{\Gamma(x)^2}{\Gamma(2x)}
=1\!+\!2(\log 2) x\! +\!{\cal O} (x^2) \ , \;\;\;\;\;\;\;
\label{linearex}
\end{eqnarray}
which was obtained \cite{KMM2} by an exact evaluation of the
path-integral with a linear tachyon profile $T = a + u_\mu X^\mu$.
Therefore, these different actions (\ref{tacbsft}), (\ref{sigmaLt}) and
(\ref{linearex}) are different expansions of the unique BSFT action
which contains correct physical quantities: the tachyon mass, the
tachyon potential an so on. 

To consider the rolling tachyon solution \cite{originalroll, LNT} in the 
obtained BSFT action (\ref{tacbsft}) is interesting. We will study it in 
appendix \ref{app:rolling}. There we show that a half S-brane marginal
deformation of a conformal field theory \cite{stro} is actually a
solution of our equation of motion with an infinite number of
derivatives, preserving the energy while the pressure gradually
decreases.  

It is noteworthy that for the rolling tachyon solutions we should
consider local quantities, such as the Lagrangian density $L$ or energy
momentum tensor, instead of the integrated values such as the action  
$S=\int\! dx\; L$ \cite{originalroll, LNT}. 
This is because in the latter case the actual
action is divergent. It follows that we can not use any partial
integration or momentum conservation relations because the integration
is divergent for such configuration. Therefore the Lagrangian density
should be sensible. Indeed, our analysis above in getting the Lagrangian
has not used the partial integration or the momentum conservation
relation. 


\subsection{Gauge field two-point function}
\label{subsec-two-point gauge}

The two-point function of the massless gauge fields can be obtained in
the same manner.  
The boundary interaction for the gauge field is (\ref{gaugeint}), and 
written in component fields as 
\begin{eqnarray}
 I_{\rm B} = -i\int_{\p \Sigma}\!\!
d\tau
 \int\!\! dk \left(
a_\mu(k)\dot{X}^\mu e^{ik_\nu X^\nu}  -2f_{\mu\nu}(k)
e^{ik_\mu X^\mu}\psi^\mu\psi^\nu
\right)
\ .
\label{gaugeintc}
\end{eqnarray}
Here $a_\mu(k)$ is the momentum representation of the target space gauge
field, and $f_{\mu\nu}(k)\equiv ik_\mu a_\nu (k)-ik_\nu a_\mu(k)$ 
is that of the field strength. 
Note that the same action is used for the gauge fields in 
a BPS D-brane, a non-BPS D-brane and a brane-anti-brane;
the following consideration is applicable to any D-brane system.
Let us define the renormalized gauge field for the boundary action to
be well-defined, 
\begin{eqnarray}
 I_{\rm B} = -i\int_{\p \Sigma}\!\!
d\tau
 \int\!\! dk \left(
a_{{\rm R}\mu}(k)\dot{X}^\mu :e^{ik_\nu X^\nu}\!:  
-2f_{{\rm R}\mu\nu}(k)
:e^{ik_\mu X^\mu}\!:\; \psi^\mu\psi^\nu
\right)
\ ,
\label{bcgauge}
\end{eqnarray}
where 
\begin{eqnarray}
a_\mu =  a_{{\rm R}\mu} \exp\left[2(\log\epsilon)k_\mu k^\mu \right]
\ ,
\quad 
A_{\mu}(X)
= \exp\left[-2(\log\epsilon)\p_\mu \p^\mu \right]
A_{{\rm R}\mu}(x) \ . 
\label{rengauge}
\end{eqnarray}
Making the operator $\dot{X}^\mu$ normal-ordered with $:e^{ik\cdot X}:$ 
may produce an additional term, 
\begin{eqnarray}
&& \int\! d\tau\; \dot{X}^\mu : e^{ik\cdot \hat{X}} :
- \int\! d\tau : \dot{X}^\mu  e^{ik\cdot \hat{X}} :
\nonumber \\
&&\hspace{10mm}=
 \int\! d\tau d\tau_1 d\tau_2 
\langle X^\rho(\tau_1) X^\sigma(\tau_2)\rangle 
\frac{\p}{\p\tau} \delta(\tau_1\!-\!\tau)
\delta(\tau_2\!-\!\tau)
\eta^{\mu\rho} ik^\sigma 
: e^{ik\cdot \hat{X}} : \ .
\label{self-cont}
\end{eqnarray}
However, this last expression vanishes
with careful treatment with the regularized propagator (\ref{regpro}).
The fermion self-contraction 
$\psi^\mu \psi^\nu-:\psi^\mu \psi^\nu:$ vanishes by the same
reason.\footnote{When there is a background constant field strength,
these are non-vanishing. See appendix \ref{app:constb}.
} 
So the well-defined boundary interaction for the renormalized
gauge field (\ref{rengauge}) is given by (\ref{bcgauge}). 

Expanding (\ref{rengauge}) to the leading nontrivial order in $k$, we
obtain 
\begin{eqnarray}
 a_\rho (k) = a_{{\rm R}\rho}(k) + 
2k^2 (\log \epsilon)a_{{\rm R}\rho}(k) \ .
\end{eqnarray}
Field redefinitions of $a_\mu$ which is of the form of a total
derivative in the boundary action is still allowed, thus we may add
a term to get
\begin{eqnarray}
 a_\rho (k) &=& a_{{\rm R}\rho}(k) + 
2k^2 (\log \epsilon)a_{{\rm R}\rho}(k) 
- 2k^\nu k_\rho (\log \epsilon)a_{{\rm R}\nu}(k) 
\nonumber \\
&=& a_{{\rm R}\rho}(k) - 
2ik^\nu (\log \epsilon)f_{{\rm R}\nu\rho}(k) \ .
\end{eqnarray}
This is the form which has been often used in the $\sigma$ model
approach, see \cite{TTU}. So our renormalization is the same as that of
the $\sigma$  model.  

We would like to evaluate the two-point function of 
the boundary coupling (\ref{bcgauge}), 
$\int dx  \langle I_{\rm B} I_{\rm B}\rangle$, 
with use of the boundary propagators (\ref{prop1}).
A straightforward calculation shows
\begin{eqnarray}
&&\langle I_{\rm B} I_{\rm B}\rangle/2\pi
\nn\\
&&
=  \int\!\! dk_1 dk_2\;
a_{{\rm R}\mu}(k_1)a_{{\rm R}\nu} (k_2)\!\!
\int\!\! d\tau\!
\left(
 \frac{\eta^{\mu\nu}}{\sin^2\frac{\tau}{2}}
\!+\!4 k_2^\mu k_1^\nu\cot^2\frac{\tau}{2}
\right)
\left\langle 
:e^{ik_1X}(\tau): :e^{ik_2X}(0):
\right\rangle
\nonumber\\
&&
\hspace{2mm}+ 
\int\!\! dk_1 dk_2\;
f_{{\rm R}\mu\nu}(k_1) f_{{\rm R}\rho\sigma}(k_2)\!\!
\int \!\!d\tau\;
\frac{
\eta^{\mu\rho}\eta^{\nu\sigma}\!-\!\eta^{\mu\sigma}\eta^{\nu\rho}
}
{2\sin^2\frac{\tau}{2}}
\left\langle 
:e^{ik_1X}(\tau): :e^{ik_2X}(0):
\right\rangle \ .
\nn
\end{eqnarray}
The remaining correlator is just (\ref{well-known}), 
so the integration over $\tau$ can be performed 
and finally gives 
\begin{eqnarray}
  \langle I_{\rm B} I_{\rm B}\rangle
&=& -4\pi^{3/2}\!\!  \int\!\! dk_1 dk_2\;
f_{{\rm R}\mu\nu}(k_1) f_{\rm R}^{\mu\nu}(k_2)
(1\!-\!4k_1\cdot k_2)\frac{\Gamma(2k_1\cdot k_2\!-\!1/2)}
{\Gamma(2k_1\cdot k_2\!+\!1)}e^{i(k_1+k_2)x}
2^{4k_1\cdot k_2}
\nonumber \\
&=& 
8\pi^{3/2}\int \! \! dk_1 dk_2\;
f_{{\rm R}\mu\nu}(k_1) f_{\rm R}^{\mu\nu}(k_2)
\frac{2^{4k_1\cdot k_2}\Gamma(2k_1\cdot k_2+1/2)}
{\Gamma(2k_1\cdot k_2+1)}e^{i(k_1+k_2)x}
\nn\\
&=&
8\pi^2 F_{{\rm R}\mu\nu}(x)
\left(
1+  \frac{\pi^2}6  (\overleftarrow{\p} \cdot\overrightarrow{\p})^2   
+ {\cal O} \Bigl((\overleftarrow{\p} \cdot\overrightarrow{\p})^3\Bigr)
\right)
F_{\rm R}^{\mu\nu}(x) \ . 
\label{gaugere}
\end{eqnarray}
This incidentally reproduces the well-known fact that the derivative
correction of the form $(\p_\mu f_{\nu\rho})^2$ can have a vanishing
coefficient for an appropriate choice of the renormalization condition
in the $\sigma$ model approach. The terms higher order in $y$ 
can also be put to be zero by a field redefinition since we are dealing
with just the two-point function. The situation becomes more nontrivial
in a constant field strength background in which one can compare more
terms with the results obtained in the other approaches,
including string scattering amplitudes. For details see appendix
\ref{app:constb}.  

In the present case the gauge invariance ensures that the mass-shell
condition is just as expected, $k^2=0$.
Indeed, if we take the Lorentz gauge $k \cdot a_{\rm R} =0$ 
or add an appropriate gauge fixing term to the action (like the $\xi=1$
(Feynman or Fermi) gauge), and use a partial integration,  
we can easily show 
\begin{eqnarray}
  \langle I_{\rm B} I_{\rm B}\rangle
&=& 
8 \pi^{3/2}\int \! \! dk_1 dk_2\;
e^{i(k_1+k_2)x}
a_{{\rm R}\mu}(k_1) a_{\rm R}^{\mu}(k_2)
 Z^{(2)}_{\rm gauge} (2 k_1 \cdot k_2) \ ,
\end{eqnarray}
where the kinetic operator is defined as 
\begin{eqnarray}
 Z^{(2)}_{\rm gauge} (y)\equiv
\frac{2^{2y}\Gamma(y+1/2)}{\Gamma(y)} 
=\sqrt{\pi} \, y(1+ {\cal O} (y^2)) \ ,
\label{gaugekin}
\end{eqnarray}
which has a zero at $y=-2 k^2=0$. The kinetic operator is regular for
$y>-1/2$ while for negative $y$ there appears periodically zeros and
poles. Its similarity to the tachyon kinetic operator (\ref{Z2tac2}) is
obvious. In fact, this structure appears commonly for all the open
string excitations as we shall see below for the massive cases.
General discussions on this structure will be given in section
\ref{sec-gene}.


\subsection{Massive field two-point functions}

Now it is clear that one can follow the same procedures to get two-point
functions also for the massive excitations. Although the boundary
couplings representing the massive excitations are generically
non-renormalizable especially when seen in Taylor (derivative)
expansions of the boundary couplings, 
in the normal-ordered plane-wave basis the dimension of the boundary
couplings can be chosen to be normalizable for appropriate values of the
off-shell momenta. Then the result can be analytically continued to give 
information even around the on-shell momentum.


\subsubsection{First massive state on non-BPS brane}

We explicitly show how this mechanism works for the first massive state
in a non-BPS D9-brane. The usual quantization on the
worldsheet theory results in a single massive antisymmetric two-form
field $B_{[\mu\nu]}(x)$ as a physical spectrum. The mass squared is 
$1/2\alpha'$ and the field is subject to the constraint 
$\p^\mu B_{\mu\nu}(x)=0$. In this subsection we reproduce this result in
the BSFT. 

In writing the most general boundary couplings, we need 
a single ${\bf \Gamma}$ and two $D_\theta$ so that the coupling
represents a consistent GSO parity and the mass level:
\begin{eqnarray}
&& I_{\rm B} = \int\! d\tau d\theta \;\biggl[
D_\theta {\bf X}^\mu D_{\theta} {\bf X}^\nu B'_{\mu\nu}({\bf X}) 
{\bf \Gamma}
+ D_\theta D_\theta {\bf X}^\mu C_\mu ({\bf X}) {\bf \Gamma}
\biggr.
\nonumber \\
&& \hspace{40mm}
\biggl.
+ D_\theta  {\bf X}^\mu E_\mu ({\bf X}) D_\theta{\bf \Gamma}
+ F_\mu ({\bf X})  D_\theta D_\theta{\bf \Gamma}
\biggr] \ . 
\end{eqnarray}
We can use a partial integration to put $E_\mu=F=0$, 
noting that the total derivative vanishes,
\begin{eqnarray}
 \int\! d\tau d\theta\; D_{\theta} \left[*\right]
=0 \ .
\end{eqnarray}
The resultant Lagrangian has the following gauge symmetry:
\begin{eqnarray}
\delta B'_{\mu\nu} = \p_\mu \Lambda_\nu - \p_\nu \Lambda_\mu \ 
, \quad  \delta C_\mu = 2 \Lambda_\mu \ 
, \quad \delta {\bf \Gamma} = -D_\theta X^\nu \Lambda_\nu \ .
\end{eqnarray}
Note that here for this transformation to be a symmetry we ignored 
$C\Lambda$ and $B \Lambda$. This is justified when we use the normal
ordering for the fields and also for the gauge transformation parameter
$\Lambda$, as in (\ref{vanisht2}). 
Then, if we use the gauge invariant combination 
$B_{\mu\nu}=B'_{\mu\nu}-\frac12 (\p_\mu C_\nu-\p_\nu C_\mu)$
instead of $B'$, the field $C_\mu$ drops 
out in the action after an appropriate change 
of ${\bf \Gamma}$.
Indeed, $C_\mu$ can be trivially gauge away and this is an analogue of the  
Higgs mechanism in the massive sector. 
Then what we should consider at this level is just
\begin{eqnarray}
&& \int\! d\tau d\theta \; \bigl[
D_\theta {\bf X}^\mu D_{\theta} {\bf X}^\nu B_{\mu\nu}({\bf X}) 
{\bf \Gamma}\bigr] \ .
\end{eqnarray}
Note that because ${\bf \Gamma}$ transforms by the gauge
transformation, the space-time fields which correspond to the GSO odd
sector, for example the tachyon, should appear in the gauge
transformation law for fields in the GSO even sector. However, these
fields come into the transformation as products with $\Lambda$,
therefore it vanishes for appropriate momentum region (as in
(\ref{vanisht2})) and we can neglect this mixing effect in our analytic
continuation method at least in the two-point functions.

Let us consider the two-point function of the resultant coupling $B$. 
We decompose the fields into component fields with use of
(\ref{xdecomp}), (\ref{supersd}) and  (\ref{bfer}). 
Then, after integrating out the auxiliary field $F$, we obtain the
following expression for the boundary coupling of the field $B(x)$ : 
\begin{eqnarray}
 I_{\rm B} = 2i\int d\tau 
\left[
\left(
2\dot{X}^\mu \psi^\nu B_{\mu\nu}(X)
-4\psi^\mu \psi^\nu \psi^\rho \p_\rho B_{\mu\nu}(X) 
\right)\eta
\right] \ .
\end{eqnarray}
Our regularization principle is to use the normal ordering for the
Fourier transform of the fields, 
\begin{eqnarray}
 B_{\mu\nu}(X) \equiv \int\! dk\; b_{{\rm R}\mu\nu} (k) :e^{ik\cdot X}: 
\ .
\end{eqnarray}
A straightforward calculation shows that
\begin{eqnarray}
&&\langle I_{\rm B} I_{\rm B} \rangle
= \int\! dkd\widetilde{k}\; e^{i(k + \widetilde{k})\cdot x}
32 \pi^{3/2} 2^{4k\cdot \widetilde{k}}
\frac{\Gamma(2k\cdot\widetilde{k})}{\Gamma(2k\cdot\widetilde{k}+1/2)}
\nonumber\\
&&\hspace{20mm}
\times
\left[
(2k\cdot \widetilde{k}-1/2)
b_{{\rm R}\mu\nu}(k) b_{\rm R}^{\mu\nu}(\widetilde{k})
-4 k^\mu b_{{\rm R}\mu\nu}(k) \widetilde{k}_\rho b_{\rm R}^{\rho\nu}
(\widetilde{k})
\right] \ . 
\label{kineb}
\end{eqnarray}
The equations of motion are solved by 
\begin{eqnarray}
 (2k^2+1/2) b_{{\rm R}\mu\nu}(k)=0, \quad 
k^\mu b_{{\rm R}\mu\nu}(k)=0 \ .
\end{eqnarray}
The first equation is the mass-shell condition $k^2=-1/4$, while the
latter equation is a constraint for the $b_{{\rm R}\mu\nu}$ field. 
These are identical with the worldsheet derivation of the spectrum, thus 
we have confirmed that our analytic continuation method in the BSFT 
provides a consistent on-shell information even for massive fields.
The number of the physical degrees of freedom is
\begin{eqnarray}
 \frac12 d(d-1) - (d-1) = \frac{1}2 (d-1)(d-2).
\end{eqnarray}
Note that the constraints give $-(d-1)$ since 
a constraint $k^\mu b_{{\rm R}\mu 0} $ does not have  $k_0$ component, 
i.e.~a time derivative, and $k^\nu (k^\mu b_{{\rm R}\mu\nu}(k))=0$
trivially. So, one of the constraint is not a dynamical one.

The kinetic term (\ref{kineb}) has a pole at
$k^2=0$. Therefore it is not well-defined at the zero momentum. 
This would be related to the fact that the massive modes are
non-renormalizable in the Taylor (derivative) expansion of the boundary
coupling, since the Taylor expansion is by definition around the zero
momentum.  


\subsubsection{First massive state on BPS D-brane}
\label{subsub24}

Let us consider the first massive state on a BPS D-brane. 
Boundary couplings at this level are composed of three
super-derivatives, so we may write down the general couplings as
\begin{eqnarray}
I_{\rm B} = \int \!\!d\tau d\theta \left[
D_\theta {\bf X}^\mu D_\theta {\bf X}^\nu D_\theta 
{\bf X}^\rho V_{\mu\nu\rho}({\bf X})
+ 
D_\theta^2 {\bf X}^\mu D_\theta {\bf X}^\nu 
W_{\mu\nu}({\bf X})
+ 
D_\theta^3 {\bf X}^\mu S_{\mu}({\bf X})
\right] \ . \;\;\;\;
\label{massbpss}
\end{eqnarray}
Due to the fermionic nature of $D_\theta {\bf X}^\mu$, the indices of 
the field $V_{\mu\nu\rho}$ are totally-antisymmetric. 
This system has the following two gauge symmetries,
\begin{eqnarray}
&& \mbox{(a)} \qquad \delta V_{\mu\nu\rho} = \frac16
\p_{[\rho} \Lambda_{\mu\nu]}({\bf X}) \ , \quad
\delta W_{\mu\nu} = \Lambda_{\mu\nu}({\bf X}) \ , \\
&& \mbox{(b)} \qquad 
\delta W_{\mu\nu} = \p_\nu \Lambda_\mu({\bf X}) \ , \quad 
\delta  S_\mu = \Lambda_\mu ({\bf X}) \ .
\end{eqnarray}
Since the indices of the field $V$ are antisymmetric, 
$\Lambda_{\mu\nu}$ in (a) is also antisymmetric in its indices. 
Thus we can gauge away the antisymmetric part of the tensor field
$W_{\mu\nu}$.  Using the second gauge symmetry (b), we may gauge away 
the field $S_\mu$. 
Remaining fields are the antisymmetric $V_{\mu\nu\rho}$ and the
symmetric part of $W_{\mu\nu}$.

In this subsection, we concentrate on the symmetric field $W_{\mu\nu}$
and shall derive its mass-shell condition by computing its two-point
function in the BSFT. (It is easy to find that there is no mixing term
among $V$ and $W$ at the level of two-point functions, due to their
symmetry property on the indices.)  
The result should recover the mass-shell condition
obtained by the old covariant quantization technique for open
superstrings,  
\begin{eqnarray}
 k^2 = -\frac12 \ , \quad 
\p^\mu W_{\mu\nu}(x)=0 \ , \quad W_{\mu}^{\;\mu}(x) =0 \ .
\label{ocqmassbps}
\end{eqnarray}
That is, the on-shell degrees of freedom of the symmetric tensor field
$W_{\mu\nu}$ are traceless and transverse to the momentum, and their
mass squared is $1/\alpha'$.

The component expression of the boundary coupling is obtained from 
the superfield expression (\ref{massbpss}) as
\begin{eqnarray}
 I_{\rm B} = \int \!\!d\tau 
\left[
4 \dot{X}^\mu \psi^\nu \psi^\rho \p_\rho W_{\mu\nu}(X)
+ (\dot{X}^\mu \dot{X}^\nu -4 \dot{\psi}^\mu \psi^\nu)W_{\mu\nu}(X)
\right] \ .
\end{eqnarray}
We expand the field in the plane-wave basis as before,
\begin{eqnarray}
 W_{\mu\nu}(X) = \int\! dk \; w_{\mu\nu}(k) e^{ik\cdot X}
= \int\! dk \; w_{{\rm R}\mu\nu}(k) :e^{ik\cdot X}: \ .
\end{eqnarray}
The renormalization due to the 
normal ordering for the plane wave basis was done with the regularized
propagator in the same manner, 
\begin{eqnarray}
w_{\mu\nu}(k)=e^{2k^2\log \epsilon} w_{{\rm R}\mu\nu}(k) \ .
\end{eqnarray}
However, necessary renormalization is not only this, in contrast to the
case of the massless states. Contractions appearing in 
a part of the boundary couplings generate a finite additional term,
since 
\begin{eqnarray}
\dot{X}^\mu \dot{X}^\nu -4 \dot{\psi}^\mu \psi^\nu
&=& :\dot{X}^\mu \dot{X}^\nu: - 4 :\dot{\psi}^\mu \psi^\nu:
+ \eta^{\mu\nu}
\left(
\frac{1}{\sinh^2(\epsilon/2)}
- \frac{\cosh(\epsilon/2)}{\sinh^2(\epsilon/2)}
\right)
\nonumber 
\\
&\rightarrow & :\dot{X}^\mu \dot{X}^\nu: - 4 :\dot{\psi}^\mu \psi^\nu:
-\frac12 \eta^{\mu\nu}
\qquad (\mbox{as} \; \epsilon \rightarrow 0)
\end{eqnarray}
The cancellation of the divergences here is due to the
supersymmetry,\footnote{Strictly speaking, the boundary condition of the
NS sector we are considering breaks the supersymmetry
\cite{Andreev-Tseytlin}.} 
but there remains a finite constant.
Because the contractions $\langle \dot{X}X\rangle$ is vanishing, 
there appears no additional term.
We obtain a well-defined boundary couplings with normal ordered
operators, 
\begin{eqnarray}
&& I_{\rm B} = \int\!\! d\tau\!\int \!\! dk \;
w_{{\rm R}\mu\nu}(k)
\biggl[
4ik_\rho :\dot{X}^\mu \psi^\nu \psi^\rho e^{ik\cdot X}\!:
\biggr.
\nonumber \\
&&\hspace{30mm}
\left.
+ :\dot{X}^\mu \dot{X}^\nu e^{ik\cdot X}\!: 
-4 :\dot{\psi}^\mu \psi^\nu e^{ik\cdot X}\!:
-\frac12 \eta^{\mu\nu}:e^{ik\cdot X}\!:
\right] \ .
\end{eqnarray}
The presence of the last term is quite unpleasant: this results in the
non-vanishing one-point function, since 
\begin{eqnarray}
 \langle I_{\rm B} \rangle = \int \!\! dk \;
  w_{{\rm R}\mu}^{\;\;\;\;\;\mu}(k)\; e^{ikx} \ .
\end{eqnarray}
The non-zero one-point function immediately means that the vacuum we
have chosen is not really a vacuum. However, it is quite unlikely  
that zero vacuum expectation value for all the excitations is not 
a consistent open string vacuum. Then what is wrong\footnote{In
\cite{Frolov}, a possibility of cancellation with a higher level boundary
term was discussed, but it may not help the cancellation of one-point
functions of the whole coupling space.}  
with this? Our standpoint on this point in this paper 
is that the one-point function
should be put to zero from the first place --- there are several natural 
reasons to believe in this prescription. First, if we closely
look at the old covariant quantization, the traceless condition
$w_{{\rm R}\mu}^{\;\;\;\;\;\mu}=0$  appears as a
physical state condition stemming from a supersymmetry generator
$G_{3/2}$ which does not include Lapracian, and this suggests that the
traceless condition does not come out as a consequence of the equation
of motion of the BSFT.  
Secondly, let us consider a corresponding procedure in the
$\beta$-function method in string $\sigma$ models. In that method, 
one regards a divergence coming from Wick 
contractions of higher operators as an equation of
motion for the coefficient fields of the remaining operators. 
In other words, this is a renormalization in the boundary
theory. However, in the present 
case, there is no coefficient field of just $:e^{ik\cdot X}:$ to
renormalize the possibly divergent quantity. That is, the coefficient of  
$:e^{ik\cdot X}:$ should be put to be zero in the $\beta$-function
method, rather than to be renormalized. 
In fact, correspondingly to this observation, there is no way to
write the $:e^{ik\cdot X}:$ term in a supersymmetric manner,
so the term of our concern in the boundary coupling 
would cause a problem unless put to be zero.

Thus  we proceed with assuming the traceless condition
$w_{{\rm R}\mu}^{\;\;\;\;\;\mu}=0$. 
The two-point function is calculated as 
\begin{eqnarray}
\langle I_{\rm B} I_{\rm B}\rangle 
&=&
\int\! dk d\tilde{k}\; e^{i(k+\tilde{k})\cdot x}
w_{{\rm R}\mu\nu}(k) w_{{\rm R}\alpha\beta}(\tilde{k})
\pi\sqrt{\pi} 
2^{4k\cdot \tilde{k}}
\frac{\Gamma(2k\cdot \tilde{k}-1/2)}
{\Gamma(2k\cdot \tilde{k}+1)}
\nonumber \\
&&\times \left[
-8 \eta^{\mu\alpha} \eta^{\nu\beta}
(2k\cdot \tilde{k})(2k\cdot \tilde{k}-1)
-32 \tilde{k}^\mu \tilde{k}^\nu k^\alpha k^\beta 
+32 \tilde{k}^\mu k^\alpha \eta^{\nu\beta}
(2k\cdot \tilde{k})
\right.
\nonumber \\
&& \left. \hspace{10mm}
+ \eta^{\mu\nu} \eta^{\alpha\beta} 
(2k\cdot \tilde{k}-1/2)
+ 4(\tilde{k}^\mu \tilde{k}^\nu\eta^{\alpha\beta}
+ \eta^{\mu\nu} k^\alpha k^\beta)
\right] \ .
\end{eqnarray}
The equations of motion follows as
\begin{eqnarray}
&&
\frac{\Gamma(-2k^2-1/2)}
{\Gamma(-2k^2+1)}
\biggl[
-8 \eta^{\mu\alpha} \eta^{\nu\beta}
(-2k^2)(-2k^2-1)
-32 k^\mu k^\nu k^\alpha k^\beta 
-32 k^\mu k^\alpha \eta^{\nu\beta}(-2k^2)
\biggr.
\nonumber \\
&& \biggl.
\hspace{20mm}+ \eta^{\mu\nu} \eta^{\alpha\beta} 
(-2k^2-1/2)
+ 4(k^\mu k^\nu\eta^{\alpha\beta}
+ \eta^{\mu\nu} k^\alpha k^\beta)
\biggr] w_{{\rm R}\mu\nu}(k)=0 \ .
\label{eqmas}
\end{eqnarray}
Trivial solutions to this equation are
\begin{eqnarray}
2k^2 = 1,2,3,4, \cdots \ .
\label{trivia}
\end{eqnarray}
We look for nontrivial solutions of the equation (\ref{eqmas}).
We first obtain two independent scalar equations by multiplying 
$k^\alpha k^\beta$ or $\eta^{\alpha\beta}$ on (\ref{eqmas}), 
\begin{eqnarray}
\frac{\Gamma(-2k^2-1/2)}{\Gamma(-2k^2)}
k^\alpha k^\beta w_{{\rm R}\alpha\beta} 
=0 \ , 
\quad
\frac{\Gamma(-2k^2-1/2)}{\Gamma(-2k^2+1)}
(2k^2+5/2)k^\alpha k^\beta w_{{\rm R}\alpha\beta} 
=0 \ . 
\end{eqnarray}
Here we have used the traceless condition 
$w_{{\rm R}\alpha}^{\;\alpha}=0$. With the momentum different from
the trivial solutions (\ref{trivia}), a unique solution to these
equations is  
\begin{eqnarray}
k^\alpha k^\beta w_{{\rm R}\alpha\beta} =0 \ .
\end{eqnarray}
Multiplying $k^\alpha$ on (\ref{eqmas}), we obtain 
\begin{eqnarray}
 \frac{\Gamma(-2k^2-1/2)}{\Gamma(-2k^2-1)}=0 
\qquad \mbox{or}\qquad k^\alpha w_{{\rm R}\alpha\beta} =0 \ .
\end{eqnarray}
When $k^2= -1/2$ which solves the first equation, we substitute it
back to (\ref{eqmas}) and obtain 
$k^\alpha w_{{\rm R}\alpha\beta} =0$. 
(However when $k^2 =0$ there appears no additional
constraint.)
On the other hand, when the second equation 
$k^\alpha w_{{\rm R}\alpha\beta} =0$ is satisfied,
substituting back it into (\ref{eqmas}), we obtain
$k^2=-1/2,0,1/2,1,\cdots$. So, in sum, we obtain two nontrivial
solutions: 
\begin{eqnarray}
k^\alpha w_{{\rm R}\alpha\beta} =
w_{{\rm R}\alpha}^{\;\;\;\;\alpha}=0, \quad k^2 = -\frac{1}{2} \ ,
\label{case1}
\end{eqnarray}
or
\begin{eqnarray}
k^\alpha k^\beta w_{{\rm R}\alpha\beta} =
w_{{\rm R}\alpha}^{\;\;\;\;\alpha}=0, \quad k^2 = 0 \ .
\label{case2}
\end{eqnarray}
Therefore the solutions of (\ref{eqmas}), different from the trivial
solutions (\ref{trivia}), are (\ref{case1}) or (\ref{case2}).
The first solution (\ref{case1}) which is the one with the lowest $k^2$ 
recovers the result of the old covariant quantization
(\ref{ocqmassbps}). (Note that in the previous cases with lower levels,
there similarly appears the additional zeros in the kinetic functions,
and the true mass-shell conditions reproducing correctly the worldsheet
spectra is the one with the lowest $k^2$.)


\section{Interaction in BSFT --- three-point function}
\label{sec-3}
\setcounter{footnote}{0}


Although the two-point functions themselves exhibit interesting
structures, for our BSFT to be a field theory, study of nontrivial
interactions is indispensable. In this section we compute a three-point
function explicitly. We consider only the gauge fields and the tachyon
fields for simplicity. For these fields, the three-point function
appears in a brane-anti-brane, while there is 
no three-point interaction for a BPS or a non-BPS D-brane. This follows
from the GSO parity and the symmetry $\tau \leftrightarrow -\tau$ 
in the partition function. 
The simplest possibility which one might expect in the brane-anti-brane
is the $A_\mu^{(-)}T\overline{T}$ three-point function, since it 
might arise as a part of the covariant derivative of the complex tachyon 
kinetic term $D_\mu T D^\mu \overline{T}$. In this section, we shall see
this in detail. 

As for the boundary couplings of tachyon and massless gauge fields on
the brane-anti-brane, we follow the notation of \cite{TTU}.
The boundary interaction terms written in component fields after
integration of the auxiliary fields are 
\begin{eqnarray}
&& I_{\rm B} = \int_{-\pi}^\pi d\tau 
\left[
\frac{i}{2}[\overline{\eta}, \eta] \dot{X}^\mu A_{\mu}^{(-)}(X)
-i[\overline{\eta}, \eta] \psi^\mu \psi^\nu F_{\mu\nu}^{(-)}(X)
+ i \sqrt{\frac{2}{\pi}}\overline{\eta}\psi^\mu D_{\mu}T(X)
\right.
\nn\\
&&\left.
- i \sqrt{\frac{2}{\pi}}\psi^\mu \eta D_{\mu}
\overline{T}(X)
-\frac{1}{2\pi} \overline{T}(X)T(X) 
+ \frac{i}{2 }\dot{X}^\mu A_{\mu}^{(+)}(X)
-i \psi^\mu \psi^\nu F_{\mu\nu}^{(+)}(X)
\right] \ .
\label{bddb}
\end{eqnarray}
The gauge fields $A^{(\pm)}$ are (plus or minus) linear combinations of
the $U(1)$ gauge fields living on two D-branes, 
$A^{(\pm)}\equiv A_\mu^{(1)}\pm A_\mu^{(2)}$. The complex tachyon
field is charged under only the gauge group of $A^{(-)}$, and the
covariant derivative is defined as $D_\mu T = \p_\mu T - i
A_\mu^{(-)}T$. 
We define the operator fields in terms of the
normal-ordered Fourier transform as before (in the following, for
simplicity  we omit the subscript ``R'' which denotes the renormalized
fields),
\begin{eqnarray}
&& T(X) = \int \!\!dk\; t(k) :e^{ik\cdot X}\!: \ , 
\quad
\tb(X) = \int\!\! dk\; \overline{t}(k) :e^{ik\cdot X}\!: \ , 
\\
&&
A_\mu(X) = \int\!\! dk\; a_\mu (k) 
:e^{ik\cdot X}\!: \ , 
\quad 
F_{\mu\nu}(X) = \int\!\! dk\; f_{\mu\nu}(k) :e^{ik\cdot X}\!: \ .
\end{eqnarray}
Note that we defined $\overline{t}(k) = t^*(-k)$. 
With this renormalized fields, for example the term $T\overline{T}$
in (\ref{bddb}) vanishes in an
appropriate region of the momenta, by the same reason as  
(\ref{vanisht2}) in the non-BPS case. When the momenta are chosen 
appropriately, any operator product at the same worldsheet point
always vanishes, thus the covariant derivatives appearing in
(\ref{bddb}) can be replaced by just the ordinary derivative, 
\begin{eqnarray}
 D_{\mu} T(X) = \p_\mu T(X) \ , 
\quad 
 D_{\mu} \overline{T}(X) = \p_\mu \overline{T}(X) \ .
\end{eqnarray}
So their Fourier transforms are
\begin{eqnarray}
 D_{\mu} T(x) = \int\! dk\; e^{ik\cdot x} ik_\mu \;t(k) \ , 
\quad
 D_{\mu} \tb(x) = \int\! dk\; e^{ik\cdot x} ik_\mu \;\overline{t}(k)
\ .
\label{normalD}
\end{eqnarray}
The propagator for $\eta$ and $\overline{\eta}$ is analogous to
(\ref{prop2}), 
\begin{eqnarray}
 \left\langle\eta(\tau) \overline{\eta}(\tau')\right\rangle
= \frac{1}{2}\epsilon(\tau-\tau') = \frac{1}{2}
\frac{\sin(\tau/2)}{|\sin(\tau/2)|} \ . 
\end{eqnarray}

The computation of the complex tachyon two-point function goes precisely 
as before, to give
\begin{eqnarray}
 \langle I_{\rm B}(T)I_{\rm B}(\overline{T})\rangle
&=&
\int\! d\tau_1 d \tau_2\;
\left\langle 
\left(i \sqrt{\frac{2}{\pi}}\right)
:\overline{\eta} \psi^\mu D_{\mu}T(\tau_1):
\left(-i \sqrt{\frac{2}{\pi}}\right)
:\psi^\nu \eta D_{\nu}\tb(\tau_2):
\right\rangle
\nonumber \\
&=&
-\int\! dk_1 dk_2\; e^{ik_1 \cdot x + ik_2 \cdot x}
 t(k_1)\overline{t}(k_2)
2^{4k_1 \cdot k_2} \sqrt{\pi} \frac{\Gamma(2k_1 \cdot k_2+1 )}
{\Gamma(2k_1 \cdot k_2 + 1/2)} \ .
\end{eqnarray}
The kinetic function is $Z^{(2)}(y)$ (\ref{Z2tac2}). Expansion for small
momenta, $y\sim 0$, coincides with the action obtained in \cite{TTU}.


\subsection{Three-point function in brane-anti-brane}

It is easy to see that the following three-point functions vanish
because some of the fermionic boundary operators do not
have their counterpart to be contracted in Wick contractions.
\begin{eqnarray}
&& TTT= TT\tb= T\tb\tb = 
 TTA^{(+)} =\tb \tb A^{(+)} 
= TTA^{(-)} =\tb \tb A^{(-)} 
= TA^{(+)}A^{(+)} 
\nn\\
&&
= TA^{(+)}A^{(-)} = TA^{(-)}A^{(-)} 
=\tb A^{(+)}A^{(+)} = \tb A^{(+)}A^{(-)} = \tb A^{(-)}A^{(-)} =0 \ .
\nn
\end{eqnarray}
Most of other three-point functions vanish due to the symmetry
$\tau \rightarrow -\tau$ :
\begin{eqnarray}
 T \tb A^{(+)} 
=A^{(+)}A^{(+)}A^{(+)} 
=A^{(+)}A^{(+)}A^{(-)} 
=A^{(+)}A^{(-)}A^{(-)} =A^{(-)}A^{(-)}A^{(-)} 
=0
\nn
\end{eqnarray}
except the single one, $T\tb A^{(-)}$. This term is expected, as
mentioned earlier. 
To evaluate this $T\tb A^{(-)}$, 
first we compute the contribution from the $F^{(-)}$ term in
(\ref{bddb}). 
\begin{eqnarray}
\lefteqn{
\int\!\! d\tau_1 d \tau_2d\tau_3
\left\langle \!
i \sqrt{\frac{2}{\pi}}\!
:\!\overline{\eta} \psi^\mu D_{\mu}T(\tau_1)\!:\!\!
\left(-i \sqrt{\frac{2}{\pi}}\right)\!\! 
:\!\psi^\nu \eta D_{\nu}\tb(\tau_2)\!:
(-i)\!
:\![\overline{\eta}, \eta] \psi^\rho \psi^\sigma 
F_{\rho\sigma}^{(-)}(\tau_3)\!:
\!\right\rangle
}
\nn\\
&&
=\frac{-2i}{\pi}\int\! dk_1 dk_2 dk_3 e^{i(k_1 + k_2 + k_3)\cdot x}
i(k_1)_\mu t(k_1) i(k_2)_\nu \overline{t}(k_2) f_{\rho\sigma}(k_3)
\nn\\
&& \quad \times
\int\! d\tau_1 d \tau_2d\tau_3
\left\langle 
:\overline{\eta} \psi^\mu e^{ik_1 \cdot \hat{X}}(\tau_1):
:\psi^\nu \eta e^{ik_2 \cdot \hat{X}}(\tau_2):
:[\overline{\eta}, \eta] \psi^\rho \psi^\sigma 
e^{ik_3\cdot \hat{X}}(\tau_3):
\right\rangle
\nn\\
&&
= \frac{-i}{4\pi}\!\!
\int\!\! dk_1 dk_2 dk_3 
e^{i(k_1 + k_2 + k_3)\cdot x}
2^{4(k_1\cdot k_2 + k+2 \cdot k_3 + k_3 \cdot k_1)}
(k_1^\rho k_2^\sigma - k_2^\rho k_1^\sigma)
t(k_1)\overline{t}(k_2) f_{\rho\sigma}(k_3)
\nn\\
&& \quad \times
\int \!d\tau_1 d\tau_2 d\tau_3
\left|\sin\frac{\tau_{12}}{2}\right|^{4k_1\cdot k_2}
\left|\sin\frac{\tau_{23}}{2}\right|^{4k_2\cdot k_3-1}
\left|\sin\frac{\tau_{13}}{2}\right|^{4k_1\cdot k_3-1}
\nn\\
&&
= \frac{-i}{\pi} (2\pi)^3\!\!
\int\!\! dk_1 dk_2 dk_3 
e^{i(k_1 + k_2 + k_3)\cdot x}
(k_1^\rho k_2^\sigma - k_2^\rho k_1^\sigma)
t(k_1)\overline{t}(k_2) f_{\rho\sigma}(k_3)
\nn\\
&&
 \quad \times I(2k_1 \cdot k_2 , 2k_2\cdot k_3 -1/2, 
2k_1 k_3 - 1/2) \ . 
\label{fttb}
\end{eqnarray}
The integral $I$ was obtained in \cite{Coletti},
\begin{eqnarray}
\lefteqn{
I(a_1,a_2,a_3) 
\equiv \int_0^{2\pi} \frac{d\tau_1 d\tau_2 d\tau_3}{(2\pi)^3}
\left|2\sin\frac{\tau_{12}}{2}\right|^{2a_1}
\left|2\sin\frac{\tau_{23}}{2}\right|^{2a_2}
\left|2\sin\frac{\tau_{13}}{2}\right|^{2a_3}
}
\nn\\
&&
=
\frac{\Gamma(a_1 + a_2 + a_3)\Gamma(1+2a_1)\Gamma(1+2a_2)\Gamma(1+2a_3)}
{\Gamma(1+a_1)\Gamma(1+a_2)\Gamma(1+a_3)
\Gamma(1+a_1+ a_2)\Gamma(1+a_1+a_3)\Gamma(1+a_2+a_3)}
\nn\\
&&
=
\frac{2^{2(a_1+a_2+a_3)}}{\sqrt{\pi}^3}
\frac{\Gamma(a_1 + a_2 + a_3)
\Gamma(a_1+1/2)\Gamma(a_2+1/2)\Gamma(a_3+1/2)}
{\Gamma(1+a_1+ a_2)\Gamma(1+a_1+a_3)\Gamma(1+a_2+a_3)} \ .
\end{eqnarray}
The term coming from $A^{(-)}$ in (\ref{bddb}) 
can be evaluated in the same manner,
\begin{eqnarray}
\lefteqn{\hspace{-5mm}
\int\! d\tau_1 d \tau_2d\tau_3
\left\langle 
\!
i \sqrt{\frac{2}{\pi}}\!
:\!\overline{\eta} \psi^\mu D_{\mu}T(\tau_1)\!:
\!\left(-i \sqrt{\frac{2}{\pi}}\right)\!
:\!\psi^\nu \eta D_{\nu}\tb(\tau_2)\!:
\!\frac{i}{2}\!
:\![\overline{\eta}, \eta] \dot{X}^\rho A_{\rho}^{(-)}(\tau_3)\!:
\right\rangle
}
\nn\\
&&
=\frac{-i}{\pi}\int\! dk_1 dk_2 dk_3 e^{i(k_1 + k_2 + k_3)\cdot x}
i(k_1)_\mu t(k_1) i(k_2)_\nu \overline{t}(k_2) a_{\rho}(k_3)
\nn\\
&& \quad \times
\int\! d\tau_1 d \tau_2d\tau_3
\left\langle 
:\overline{\eta} \psi^\mu e^{ik_1 \cdot \hat{X}}(\tau_1):
:\psi^\nu \eta e^{ik_2 \cdot \hat{X}}(\tau_2):
:[\overline{\eta}, \eta] \dot{X}^\rho
e^{ik_3\cdot \hat{X}}(\tau_3):
\right\rangle
\nn\\
&&
=\frac{1}{2\pi}\int\! dk_1 dk_2 dk_3 e^{i(k_1 + k_2 + k_3)\cdot x}
k_1\cdot k_2 t(k_1)  \overline{t}(k_2) a_{\rho}(k_3)
2^{4(k_1\cdot k_2 + k_2 \cdot k_3 + k_1 \cdot k_3)}
\nn\\
&& \quad \times
\int\! d\tau_1 d \tau_2d\tau_3
\left|\sin\frac{\tau_{12}}{2}\right|^{4k_1\cdot k_2-2}
\left|\sin\frac{\tau_{23}}{2}\right|^{4k_2\cdot k_3-1}
\left|\sin\frac{\tau_{13}}{2}\right|^{4k_1\cdot k_3-1}
\nn\\
&&
\quad \times \left(
k_1^\rho \sin\frac{\tau_{23}}{2} 
\cos\frac{\tau_{13}}{2}\sin\frac{\tau_{12}}{2}
+k_2^\rho \sin\frac{\tau_{13}}{2} 
\cos\frac{\tau_{23}}{2}\sin\frac{\tau_{12}}{2}
\right) \ .
\end{eqnarray}
To evaluate the last part, we use the following identity
\begin{eqnarray}
\lefteqn{ \sin A \sin B \cos C }
\nn\\
&&= \frac14
\left[
-\cos(A\!+\!B\!-\!C)+ \cos(-\!A\!+\!B\!+\!C)+ \cos (A\!-\!B\!+\!C)
-\cos (A\!+\!B\!+\!C)
\right] \ . 
\nn
\end{eqnarray}
Then the last integral can be performed to give
\begin{eqnarray}
\lefteqn{
2(2\pi)^2\int\! dk_1 dk_2 dk_3\; e^{i(k_1 + k_2 + k_3)\cdot x}
k_1\cdot k_2 t(k_1)  \overline{t}(k_2) 
}
\nn\\
&&
\hspace{30mm}\times 
\bigl[k_1\cdot a(k_3)(-I_1-I_2+I_3)
- (k_1 \leftrightarrow k_2)
\bigr],
\label{attb}
\end{eqnarray}
where 
\begin{eqnarray}
 I_1 \equiv
 I\left(\!\alpha, \beta\!-\!\frac12, \gamma\!-\!\frac12\right) ,
 I_2 \equiv I\left(\!\alpha\!-\!1, \beta\!+\!\frac12, 
\gamma\!-\!\frac12\right) ,
 I_3 \equiv I\left(\!\alpha\!-\!1, \beta\!-\!\frac12, \gamma\!+\!
\frac12\right) ,
\nn
\end{eqnarray}
and $\alpha \equiv 2 k_1 \cdot k_2$, $\beta \equiv 2 k_2 \cdot k_3$, 
$\gamma \equiv 2 k_1 \cdot k_3$.

Summing up (\ref{fttb}) and (\ref{attb}), 
we obtain the full three-point function, 
\begin{eqnarray}
&&
\int\! dk_1 dk_2 dk_3\; e^{i(k_1 + k_2 + k_3)\cdot x}
t(k_1)\overline{t}(k_2)
\left[
-\beta k_1 \cdot a(k_3) 
+ \gamma k_2 \cdot a(k_3) 
\right] {\cal C} \nn\\
&&
= \int\! dk_1 dk_2 dk_3\; e^{i(k_1 + k_2 + k_3)\cdot x}
t(k_1)\overline{t}(k_2)
\nn\\
&& \hspace{20mm}\times
\frac{1}{2} 
\bigl[
(\beta-\gamma) k_3 \cdot a(k_3) 
-(\beta+\gamma) (k_1-k_2) \cdot a(k_3) 
\bigr] {\cal C} \ , 
\label{tpf}
\end{eqnarray}
where
\begin{eqnarray}
 {\cal C} & \equiv &  
\frac{\sqrt{\pi} 2^{2(\alpha+\beta+\gamma+1/2)} 
\Gamma(\alpha+\beta + \gamma+1)
\Gamma(\alpha+1/2)\Gamma(\beta)\Gamma(\gamma)}
{\Gamma(\alpha+\beta+1/2)
\Gamma(\alpha+\gamma+1/2)\Gamma(\beta+\gamma+1)} 
\ .
\label{cvalue}
\end{eqnarray}
In the Lorentz gauge $k_3 \cdot a(k_3)=0$,
the three-point function becomes a rather simple form:
\begin{equation}
\int\!\! dk_1 dk_2 dk_3\; e^{i(k_1 + k_2 + k_3)\cdot x}
t(k_1)\overline{t}(k_2)
(k_2\!-\!k_1)\! \cdot\! a(k_3) 
\frac{\sqrt{\pi}}{2} 
\frac{\Gamma(\sum_i \alpha_i\!+\!\frac{1}{2}) 
\prod_i 4^{\alpha_i} \Gamma(\alpha_i) }
{\prod_{i<j} \Gamma(\alpha_i+\alpha_j)} \ ,
\end{equation}
where $\alpha_1\equiv \alpha+1/2$, 
$\alpha_2\equiv\beta$ and $\alpha_3\equiv\gamma$.


\subsection{Exact coincidence with derivative-expanded BSFT}

As noted before, the three-point function $T \overline{T}A^{(-)}$ is
expected to arise naturally as a part of the tachyon kinetic term with 
the covariant derivative, $D_\mu T D_\mu \overline{T}$. This is the
usual picture considered in \cite{TTU}, while in our case, the covariant
derivatives appearing in the boundary interaction vanish due to our
normal ordering, (\ref{normalD}). In this sense, the origins of the 
$T \overline{T}A^{(-)}$ three-point function are quite different, but we
shall see in this subsection that these two methods give the same 
$T \overline{T}A^{(-)}$ with a suitable field redefinition. In section
\ref{sec-analytic}, 
we have seen nontrivial coincidence on the tachyon potential term 
$T^2$ in different approaches. Here is another nontrivial example.

The tachyon two-point function $T\overline{T}$ and
the three-point function $T \overline{T}A^{(-)}$ 
are included in the $\alpha'$-expanded action obtained in \cite{TTU} in
a $\sigma$ model approach, 
\begin{eqnarray}
 \frac{L}{\cal T} = -T\tb
+ 8 \log 2 D_\mu T D_\mu \tb
+ 8 \gamma_0 D_\mu D_\nu T D_\mu D_\nu \tb
+ 64i (\log 2)^2 F_{\mu\nu} D_\mu T D_\nu \tb \ .
\;\;\;\;
\end{eqnarray}
Here and in the following study we omit the suffix $(-)$ for simplicity,
and neglect terms quartic (or in higher powers) in fields and also terms
of ${\cal O}({\alpha'}^3)$. Relevant terms are expanded explicitly to
give the three-point function as 
\begin{eqnarray}
&& 
D_\mu T D_\mu \tb = \p_\mu T \p_\mu \tb
-i A_\mu (T \p_\mu \tb - \p_\mu T \tb) + {\cal O}(A^2) \ ,
\\
&&
D_\mu D_\nu T D_\mu D_\nu \tb = \p_\mu \p_\nu T \p_\mu\p_\nu \tb
-2i A_\mu (\p_\nu T \p_\nu\p_\mu \tb - \p_\nu\p_\mu T \p_\nu\tb) 
\nonumber
\\
&& \qquad \qquad \qquad 
-i(\p_\nu A_\nu)(T \p_\nu \p_\nu T - \p_\nu\p_\nu T \tb)
+ {\cal O}(A^2) \ ,
\\
&&
F_{\mu\nu} D_\mu T D_\nu \tb 
= 
(\p_\nu A_\nu - \p_\nu A_\mu) \p_\mu T \p_\nu \tb
+ {\cal O}(A^2) \ .
\end{eqnarray}
We go to the momentum representation of this Lagrangian for our later
purpose. 
\begin{eqnarray}
&& S = {\cal T}(S_2 + S_3) \ , \label{ttulag}\\
&& S_2 \equiv 
\int\! dx\!\int\! dk_1 dk_2 \; e^{i(k_1+k_2)\cdot x}
t(k_1) \overline{t}(k_2)
\left(
-1+ 8 \log 2 k_1\cdot k_2 + 8 \gamma_0 (k_1 \cdot k_2)^2
\right) \ , \;\;\;\;\;
\\
&&S_3 \equiv
\int\! dx \!\int\! dk_1 dk_2 dk_3
\; e^{i(k_1 + k_2 + k_3)\cdot x} t(k_1)\overline{t}(k_2)
\nonumber \\
&& \quad \times
\left[
k_1 \cdot a(k_3)
\left(
-8\log 2 + 8 \gamma_0 (k_1 \cdot k_3 + 2 k_1 \cdot k_2)
-64 (\log 2)^2 k_2 \cdot k_3
\right)
\right.
\nonumber 
\\
&& \quad\quad\quad \left. 
+ k_2 \cdot a(k_3)
\left(
8\log 2 - 8 \gamma_0 (k_2 \cdot k_3 + 2 k_1 \cdot k_2)
+64 (\log 2)^2 k_1 \cdot k_3
\right) 
\right] \ .
\end{eqnarray}
This Lagrangian should coincide with our BSFT Lagrangian up to a
field redefinition. This redefinition should be in higher order in
fields, since we already know that our tachyon two-point function
coincide with that of \cite{TTU} without any field redefinition. 
This means that the redefinition should be of the form
$T \rightarrow T + T A$. At a glance this redefinition looks strange,
in view of that the gauge transformation on $T$ is modified.
However, it turns out that this is the case: in our normal-ordered
plane-wave basis the tachyon is actually gauge-invariant, because 
any field multiplication turns out to be vanishing as in (\ref{vanisht2}) 
and thus $e^{i\Lambda}T = T$. (This statement should be understood
except for global gauge transformations.)
Therefore, what we need as a field
redefinition is the one which makes the tachyon field gauge-invariant.
To achieve this, we consider the following form of the field
redefinition\footnote{There are other forms satisfying our requirement
of the change of the gauge transformation, but (\ref{fd1}) and
(\ref{fd2}) turn out to be the correct one.}: 
\begin{eqnarray}
&&t(k_1) \rightarrow 
t(k_1) + 
\int\! dk_3 \frac{1}{k_3 \cdot(k_1-k_3)} 
a_\mu(k_3) (k_1-k_3)^\mu 
t(k_1-k_3) + {\cal O}(a^2) \ , 
\label{fd1}\\
&&\overline{t}(k_2) \rightarrow 
\overline{t}(k_2) -
\int\! dk_3 \frac{1}{k_3 \cdot(k_2-k_3)} 
a_\mu(k_3) (k_2-k_3)^\mu 
\overline{t}(k_2-k_3) + {\cal O}(a^2) \ .
\label{fd2}
\end{eqnarray}
Substituting this redefinition to the above Lagrangian (\ref{ttulag}),
we obtain the following $T\overline{T}A^{(-)}$ terms (we have redefined
the momenta as $k_1-k_3\rightarrow k_1$ and so on  
so that all the fields have common arguments)
\begin{eqnarray}
&&
\int\! dk_1 dk_2 dk_3 \; e^{i(k_1 + k_2 + k_3)\cdot x} t(k_1)
\overline{t}(k_2)
\nonumber \\
&& \quad
\times 
\biggl[
k_1 \cdot a(k_3)
\Bigl\{
-8\log 2 + 8 \gamma_0 (k_1\cdot k_3 + 2 k_1 \cdot k_2)-64 (\log 2)^2
k_2 \cdot k_3
\Bigr.\biggr.\nonumber \\
&& \qquad\qquad \biggl.\Bigl.
+ \frac{1}{k_1 \cdot k_3}
(-1-8\log 2 (k_1 + k_3)\cdot k_2 + 8 \gamma_0 ((k_1 + k_3)\cdot k_2)^2)
\Bigr\}
\biggr.
\nonumber
\\
&&
\;\;\;\;\;\;\;\;+k_2 \cdot a(k_3)
\Bigl\{
8\log 2 - 8 \gamma_0 (k_2\cdot k_3 + 2 k_1 \cdot k_2)+64 (\log 2)^2
k_1 \cdot k_3
\Bigr.\nonumber \\
&& \qquad\qquad \biggl.\Bigl.
- \frac{1}{k_2 \cdot k_3}
(-1-8\log 2 (k_2 + k_3)\cdot k_1 + 8 \gamma_0 ((k_2 + k_3)\cdot k_1)^2)
\Bigr\}
\biggr] \ . 
\label{tture}
\end{eqnarray}

On the other hand, in terms of small momenta
we Laurent-expand ${\cal C}$ (\ref{cvalue}) in our BSFT Lagrangian 
and obtain
\begin{eqnarray}
&& {\cal C} = \frac{1}{3\beta\gamma}
\bigl[
6+12\log 4 (\alpha + \beta + \gamma)
\bigr.
\nn\\
&& \left. \quad \quad 
+12 (\log 4)^2 (\alpha + \beta + \gamma)^2
- \pi^2 ((\alpha + \beta + \gamma)^2-2\beta\gamma) 
+ {\cal O}(k^6)\right] \ .
\end{eqnarray}
Substituting this expression to the three-point function (\ref{tpf}), 
one can see exact coincidence with the one obtained by the field
redefinition of the results of \cite{TTU}, (\ref{tture}).


\subsection{Reproduction of string scattering amplitude}
\setcounter{footnote}{0}

In this subsection, we shall see that our three-point function
reproduces string scattering amplitude at the on-shell momenta.
This is a very important check for our claim that the BSFT evaluated
with the analytic continuation gives consistent perturbative string
theory information, not only the mass-shell conditions.

First we study the value of the three-point function
when all the external momenta are set to their on-shell values.
The on-shell values of the Lorentz-invariant momentum parameters are
\begin{eqnarray}
\alpha = -1/2, \quad \beta=\gamma=0 \ . 
\end{eqnarray}
We expand the momenta around their on-shell values, 
$\alpha_1 (\equiv \alpha + 1/2), \alpha_2(\equiv\beta), 
\alpha_3(\equiv \gamma) \ll 1$.
The expansion leads
\begin{eqnarray}
{\cal C} = \pi
\frac{(\alpha_1+\alpha_2)(\alpha_1+\alpha_3)}
{\alpha_1\alpha_2\alpha_3} + {\cal O}(\alpha_i) \ .
\end{eqnarray}
This term gives indefinite result for the on-shell value of the 
three-point function --- the value depends on how one takes the limit to 
the on-shell momenta. Even worse, the expression may diverge in 
some limit, for example, $\alpha_2 \sim \alpha_3 \sim \epsilon^2, 
\alpha_1\sim\epsilon, \epsilon \rightarrow 0$.
However, we can subtract a part of this term by a field redefinition of
the tachyon, to make it definite.
The redefinition for the tachyon is of the form same as that of the
previous subsection. After subtracting these indefinite terms, we arrive
at an expression which is irrelevant on how is the limit to the on-shell 
momenta.  

Let us see this in detail.
The above limiting behavior in ${\cal C}$ results in the three-point
function in the Lorentz gauge as 
\begin{eqnarray}
&&\int\! dk_1 dk_2 dk_3\; e^{i(k_1 + k_2 + k_3)\cdot x}
t(k_1)\overline{t}(k_2)
\nn\\
&& \hspace{20mm}\times
\frac{-\pi}{2} 
(k_1-k_2) \cdot a(k_3) 
\frac{(\alpha_1+\alpha_2)(\alpha_1+\alpha_3)
(\alpha_2+\alpha_3) }
{\alpha_1\alpha_2\alpha_3} + \cdots \ .
\end{eqnarray}
The indefinite part can be cast into the form
\begin{eqnarray}
\frac{(\alpha_1+\alpha_2)(\alpha_1+\alpha_3)
(\alpha_2+\alpha_3) }{\alpha_1\alpha_2\alpha_3} 
= \frac{\alpha_1+\alpha_2}{\alpha_3}
+ \frac{\alpha_1+\alpha_3}{\alpha_2}
+ \frac{\alpha_2+\alpha_3}{\alpha_1} 
+ 2 \ .
\label{decin}
\end{eqnarray}
The first term can be eliminated by a field redefinition of the form
similar to (\ref{fd1}) and the second by (\ref{fd2}), while the third
one can be eliminated by a field redefinition of the gauge field
of the form
\begin{eqnarray}
 a_\mu(k_3) \rightarrow a_\mu(k_3) + \int\! dk\;
\frac{f(k,k_3)}{4k\cdot(k_3-k)+1} t(k_3-k)\bar{t}(k)
\label{gredef}
\end{eqnarray}
where $f$ is chosen in such a way that the resulting term coming out of
the gauge kinetic term can eliminate the third term in (\ref{decin}).
(The denominator corresponds to the factor $1/\alpha_1$.)\footnote{It
is noteworthy that the same field redefinition simultaneously removes
the divergence of the three-point function at zero momenta, because the
field redefinition of the gauge field (\ref{gredef}) is regular at the
vanishing momenta.}
The last term in (\ref{decin}) is remaining and cannot
be absorbed into the field redefinition, because
any redefinition should not be singular at on-shell momenta and so for
example a redefinition $ a_\mu \rightarrow a_\mu + \int
\frac{1}{\alpha_2 + \alpha_3}a_\mu t$
is not allowed ($\alpha_2 + \alpha_3 =-2k_3^2$  is included in the
kinetic term).
Thus the on-shell three-point function reads with the last term in
(\ref{decin}) as 
\begin{eqnarray}
L_3 = 2{\cal T_{\rm D9}}N_T^2 N_A
(-\pi) \int\!\! dk_1 dk_2 dk_3\; e^{i(k_1 + k_2 + k_3)\cdot x}
t(k_1)\overline{t}(k_2)\;
(k_1-k_2)^\mu a^{(-)}_\mu(k_3) \ . \;\;\;\;\;
\label{ourr}
\end{eqnarray}
Here $2 {\cal T}_{\rm D9}$ is the tension of the brane-anti-brane,
and we have newly included the normalization factors for the tachyon and
the gauge field $N_T$ and $N_A$ respectively in the boundary coupling:
$ T(x) \rightarrow N_T T(x)$, 
$A_\mu^{(\pm)} \rightarrow N_A A_{\mu}^{(\pm)}$.  
We have put $N_T = N_A =1$ in the calculations so far, but we need these
hereafter to compare our result with a string scattering amplitude.
The normalization factors $N_{T,A}$ can be fixed by requiring 
the canonical normalization for the kinetic term (two-point functions). 
The tachyon kinetic term can be expanded as (see (\ref{coefz2}))
\begin{eqnarray}
 2 {\cal T}_{\rm D9} N_T^2 T(x) \left(\pi 
\left[
\overleftarrow{\p}\cdot\overrightarrow{\p}-\frac14
\right]
+ {\cal O}\left(
 \left[
\overleftarrow{\p}\cdot\overrightarrow{\p}-\frac14
\right]^3\right)
\right)\overline{T} \ .
\end{eqnarray}
Thus, up to a higher order field redefinition of the form 
\begin{eqnarray}
 t(k) \rightarrow 
\left(
1+ {\cal O}\left(
(k^2+1/4)^2
\right)
\right)
t(k) \ , 
\end{eqnarray}
the canonical normalization of the kinetic term 
($L=|\p_\mu T|^2-\frac14 |T|^2$) implies
\begin{eqnarray}
 2 {\cal T}_{\rm D9} N_T^2 \pi = 1 \ .
\label{tachyonnorm}
\end{eqnarray}
In the same manner, we obtain 
the gauge two-point function for the present 
brane-anti-brane case\footnote{To obtain the expression (\ref{--2}), 
we have to multiply 
$\langle :\!\overline{\eta}\eta(\tau_1)\!:
\;:\!\overline{\eta}\eta(\tau_2)\!:\rangle=1/4$
on the previous result for the BPS D-brane (\ref{gaugere}). 
The resultant normalization in (\ref{--2}) coincides with
the result of the $\sigma$ model calculation in \cite{TTU}.} 
\begin{eqnarray}
 2{\cal T}_{\rm D9}\frac12 \left\langle
I_{\rm B}(A^{(-)})I_{\rm B}(A^{(-)})
\right\rangle
= 2 {\cal T}_{\rm D9}N_A^2 \pi^2 
F_{\mu\nu}^{(-)}(x)F^{(-)\mu\nu}(x) 
\hspace{10mm}
\nonumber \\
+ \mbox{(higher derivatives)}
\label{--2}
\end{eqnarray}
with a similar expression also for $A^{(+)}_\mu$. This
gives the following normalization relation so that 
$L = \frac14 (F_{\mu\nu}^{(1)})^2 + \frac14 (F_{\mu\nu}^{(2)})^2$ 
is achieved,\footnote{We have implicitly used the same normalizations
for $A^{(+)}$ and $A^{(-)}$, because otherwise these canonical kinetic
terms for $A^{(1,2)}$ would not be achieved.}
\begin{eqnarray}
 2 {\cal T}_{\rm D9} N_A^2 \pi^2 = \frac18 \ .
\label{gaugenorm}
\end{eqnarray}
Since we know the tension of the D9-brane, the normalizations $N_T$ and
$N_A$ are completely fixed\footnote{Since the tension is of order 
$1/g_{\rm open}^2 \sim 1/g_{\rm closed}$, 
the determined normalizations $N_{T,A}$ are of order $g_{\rm open}$,
which is consistent. To determine the exact value of $N_{T,A}$
in terms of $g_{\rm open}$, we need the explicit expression for the
D-brane tension 
written by the open string coupling defined in our boundary couplings. 
This can be obtained by a computation of a one-loop amplitude of an open
string normalized in our convention and using the open-closed duality.
} from (\ref{tachyonnorm}) and (\ref{gaugenorm}), 
and thus the normalization of
the three-point function (\ref{ourr}) is determined. 

We may compare this normalized three-point function (\ref{ourr})
with the known tree-level string scattering amplitude  
\cite{Pesando}, 
\begin{eqnarray}
e (2\pi)^{10} \delta\left(\sum k_i\right)
 (k_1-k_2)\cdot \zeta^{(-)} \ , 
\label{amp}
\end{eqnarray}
where $\zeta_\mu^{(-)}$ is the polarization of the gauge field
$A_\mu^{(-)}$, and $e$ is the open string coupling defined in
\cite{Pesando}. We can deduce 
\begin{eqnarray}
 e = N_A
\label{eNA}
\end{eqnarray}
by just looking at the structure of the covariant derivatives in ours
and \cite{Pesando}. In \cite{Pesando}, the constant Wilson line was
introduced as a background which shifts the tachyon momentum as 
$k_\mu - e A^{(-)\rm b.g.}_\mu$, while in our case the change of the
normalization of the gauge field 
$A_\mu^{(-)} \rightarrow N_A A_\mu^{(-)}$ gives rise to the change of
the covariant derivative to $\p_\mu - i N_A A_\mu^{(-)}$. 
Substituting this relation (\ref{eNA}) and the tachyon normalization
(\ref{tachyonnorm}) into our three-point function (\ref{ourr}), we find
the exact reproduction of the scattering amplitude (\ref{amp}). Thus we
conclude that the BSFT three-point function (\ref{ourr}) in which the
divergent part has been subtracted by a field redefinition coincides
with the string scattering amplitude.\footnote{An amusing point is 
that a naive comparison of our boundary couplings with the string vertex
operators gives the relation same as (\ref{eNA}). The vertex operators
in the 0 picture in \cite{Pesando} were  
\begin{eqnarray}
&&
{\cal V}_T = -4 e \; c \; 
k_\mu \psi^\mu e^{ik\cdot X} \cdot \frac12(\sigma_1\pm i\sigma_2) \ , 
\label{vo1}
\\
&&
{\cal V}_A = i e \; c \;
\zeta_\mu \left(
\dot{X}^\mu - 4i \psi^\mu k_\nu \psi^\nu
\right) e^{ik\cdot X} \cdot \frac12(\sigma_0\pm \sigma_3) \ , 
\label{vo2}
\end{eqnarray}
where $c$ is the worldsheet ghost and the Pauli matrices $\sigma$'s are
Chan-Paton factors ($\sigma_0 \equiv 1_{2\times 2}$). Comparing these
with our boundary couplings by simply dropping the ghost, we obtain
\begin{eqnarray}
-N_T \sqrt{\frac{2}{\pi}}=-4 e \ , 
\quad N_A =  e  \ . 
\label{tacnorma}
\end{eqnarray}
The second relation is precisely what we learned in comparison of the
covariant derivatives (that is, the gauge transformation laws after the
field redefinition). The ratio of the relations (\ref{tacnorma}) is
consistent with the ratio of our results (\ref{tachyonnorm}) and
(\ref{gaugenorm}). Note that the normalization of the vertex operators
(\ref{vo1}) and (\ref{vo2}) was fixed in \cite{Pesando} by demanding the
unitarity (this is the reason why we had to normalize our BSFT two-point
function canonically, to compare our result with the scattering
amplitude). This suggests that our BSFT is automatically 
unitary by construction. However, in this comparison of the vertex
operators, it is not clear why we may simply drop the ghost. 
}

Note that in the bosonic case \cite{Frolov}
the tachyon three-point function with
on-shell momenta was evaluated by taking the limit symmetric under the 
momentum exchange among the three tachyons. However, with a careful
evaluation, this turns out to be unnecessary
-- one can in fact show that the three-point function given in
\cite{Coletti} is free of indefiniteness around the on-shell momenta. 
So the value of the bosonic tachyon three-point function does not depend
on how one takes the on-shell limit, and is definite without any
field-redefinition.  

Second, let us see what happens to the three-point functions when one of
the fields is set to their on-shell values. 
When the tachyon $T$ (or $\overline{T}$) is on-shell and thus 
$\alpha_1 + \alpha_3=0$ (or $\alpha_1 + \alpha_2=0$ respectively), 
the factor ${\cal C}$ vanishes and so the three-point function
disappears. When the gauge field is on-shell ($\alpha_2 + \alpha_3=0$
and $k_3 \cdot a(k_3)=0$), although ${\cal C}$ is non-vanishing, the
coefficient in front of ${\cal C}$ in (\ref{tpf}) vanishes. 
Therefore, when one of the three outer legs is on-shell, the three-point
function vanishes. 
This results in the vanishing of on-shell one-particle-reducible 
diagrams, when we treat the BSFT action as a field theory action and
perform the usual Feynman rule for getting higher point functions.
In fact, although the propagator connecting the vertices is diverging
at the on-shell momentum, it is not powerful enough to cancel the
vanishing of the on-shell vertices connected to two boundary points of
the propagator in the Feynman graph.
This is quite satisfactory since, as mentioned in the introduction,
the BSFT action $S_{\rm BSFT}=Z$ already includes the vertices which
reproduce  the on-shell tree level S-matrix \cite{Andreev-Tseytlin}, and 
one-particle-reducible Feynman graphs generated by lower vertices
should vanish.\footnote{
The paper \cite{Frolov}  considered a bosonic case in which
there are corrections containing the beta function to the relation 
$S_{\rm BSFT}=Z$. In our superstring case, there is no such correction.} 


\section{Generalities --- minimum length in string theory}
\label{sec-gene}

In this section we consider general properties of the super BSFT action.
If we look at the structure of
the worldsheet boundary integral, it is easy to notice that for massive
fields with the mass-shell condition $k^2 = -(N-1)/2\alpha'$, 
the integral is
convergent for 
\begin{equation}
2\alpha' k\cdot\tilde{k}>N \ ,
\label{reigeon}
\end{equation}
where $N$ is the oscillator level of the open string excitations. 
In our boundary couplings, $N$ is the number of the derivative
$D_\theta$. This can be seen in the the dimensions of the corresponding
operators. So, for two-point functions in BSFT, there is a region for
the momentum where the integral over the worldsheet boundary is
well-defined and finite. This is interesting in view of the fact that
usually in BSFT the massive excitations are believed not to be treatable
because they are 
non-renormalizable operators. This convergent region (\ref{reigeon}) of
the momentum can apply also for three- or more point functions, since
the divergence 
appears only when two of the vertices collide to each other. Therefore,
if any pair of the momenta satisfies (\ref{reigeon}) the integral is
finite. Then we can use the analytic continuation to obtain the
BSFT action for any momenta. 

However, there are many singularities in the momentum space.
For massive excitations, the two-point functions are
expected to have the following form: 
\begin{eqnarray}
Z^{(2)}(y) \sim \frac{2^{2y} \Gamma(y + (2-N)/2)}
{\Gamma(y + (1-N)/2)} \ ,
\end{eqnarray}
where 
$y = -\alpha' 
\overleftarrow{\p}\!\cdot\!\overrightarrow{\p} =- \alpha' k^2$.
All of our BSFT results are of this form\footnote{For a related
discussion, see \cite{Frolov}. {\bf }} --- 
the tachyon ($N=0$) $Z^{(2)}(y)$ is exactly above, 
and the gauge field ($N=1$) kinetic function takes this form 
in the Lorentz gauge $k^\mu a_\mu=0$, which is also the case for the
massive fields. From this expression, we observe that there is a 
common structure in the kinetic term of the open string excitations ---
an infinite number of poles and zeros appear in the tachyonic region
($k^2>0$) in the kinetic operator $Z^{(2)}$. For the tachyon, see
Fig.~\ref{z2fig}. Note that for $k^2>-(N-1)/2\alpha'$ the operator with
the level $N$ is irrelevant and the first pole is found at
$k^2=-(N-2)/2\alpha'$. This means that we can make an analytic
continuation of the two-point function $Z^{(2)}$ to the inside of the
irrelevant region until hitting the pole.

This quite intriguing pole/zero structure might suggest a minimum length
in space which can be observed by fundamental strings. Let us regard the
BSFT action as an off-shell generalization of the tree level S-matrix
generating effective action as in \cite{Andreev-Tseytlin}. Near the
on-shell momentum $k$, we may try to redefine the fields so that they
have canonically normalized kinetic terms. For example, the redefinition
of the gauge field should be 
$A'_\mu=(Z^{(2)}_{\rm gauge}(\alpha'\p^2))^{1/2} A_\mu$.
However, $Z^{(2)}_{\rm gauge}(y)$ (\ref{gaugekin}) has a singularity at
$y=-1/2$. Moreover, it changes the sign when we go over the singularity,
which implies that the field redefinition becomes imaginary, then is 
not allowed. Therefore we might have to restrict the region of the
momentum $k$ in order to avoid the singularity,\footnote{
Here we assume our analytic continuation is valid. We note that the
tachyon kink solution $T=u_9 X^9$, for example, has an expression 
$t(k)= i u_9 \frac{\p}{\p k}\delta(k)$ in the momentum space, which is
within the allowed momentum region even though it is an off-shell
configuration.} 
at least if we naively interpret the two-point functions, derived by
using a coordinate system of the space of the boundary couplings
(spacetime fields) suggested in BSFT. The introduction of the upper-bound
for the momentum $k^2$, somewhat like a Briroinn zone, implies that the
space becomes effectively discretized with the scale $\sqrt{\alpha'}$,
that is, the string length. One can interpret this as a spacetime
resolution, or rather to say, the minimum length in string theory. 

On the other hand, one can in principle compute the effective action
using cubic string filed theory \cite{Wittencubic} at least for the
bosonic case. This should coincide with our BSFT action upto field
redefinition ambiguities. For on-shell fields, these two are considered
to be the same as described in \cite{Coletti} for bosonic three-point
tachyons and as agrgued in \cite{Andreev-Tseytlin}. The problem is, as
mentioned above, that the field redefition used for relating 
$Z^{(2)}$ to the standard kinetic terms in the cubic string field theory 
does not exist at and beyond the singularity of $Z^{(2)}$. Furthermore,
if we remember the Witten-Shatashvili formula \cite{BSFT}
$(\delta/\delta g_i)S_{\rm BSFT} \sim \beta^i G_{ij}$ (where $\beta^i$
is the beta function for the boundary coupling (spacetime field) $g_i$, 
and $G_{ij}$ is the metric of the space of the couplings) and 
$\beta^i \sim (k^2+(N-1)/2\alpha')$, we find that the extra poles and
zeros are coming from  the metric $G_{ij}$ as was pointed out in
\cite{Coletti}. This indicates that the problem is not only for the
kinetic terms, but a more general one.

One possible resolution is that the cubic string field theory may 
also have the restriction on the momentum for some reason. However, this
is not likely because non-singular field redefinition does not relate  
interaction terms to the kinetic term at tree level. Another possibility
is that the coordinate system of the BSFT is singular at the singularity
though the physics is not singular and we should use another appropriate
coordinate system of the space of the boundary couplings beyond the
singularity. Since the fields of the BSFT is in some sense natural,
it is reasonable to expect that the singularity of the coordinate system
means some peculiar physics appearing there, like the example of the 
noncommutative soliton discussed by Sen \cite{SenNC}. Hence the
singularity of the space of the boundary couplings may reflect some kind
of the minimum length. 

Meanwhile, let us regard the BSFT action as that of a constructive
(string) field theory \cite{BSFT}, instead of taking it as an effective
action. Then in this interpretation we would have to path-integrate the
fields, which causes a serious problem for loop amplitudes. Although the
loop amplitudes of the BSFT are known to be difficult to deal with
and we do not have any definite answer to that, we would like to give a
few comments on it. The problem is that we have to perform an
integration over the loop momenta in perturbative evaluation of the loop
amplitudes. It seems that the momentum bound which we studied in this
section is an obstacle to perform the loop momentum
integration. Furthermore, on internal vertices, the momentum
conservation cannot satisfy the lower bound for the momenta. To avoid
this problem, we can decompose the propagator obtained in the BSFT as 
\begin{equation}
\frac{\Gamma(y)}{\Gamma(y+1/2)  } =\frac{1}{\sqrt{\pi} }
\sum_{n=0}^{\infty} \frac{(2n-1)!!}{(2n)!!} \frac{1}{y+n}
=\frac{1}{\pi }
\sum_{n=0}^{\infty} \frac{ \Gamma(n+1/2) }{ \Gamma(n+1) } 
\frac{1}{y+n} \ ,
\end{equation}
which is a sum of a massless propagator and infinitely many tachyonic 
propagators. (This formula is similar to the one used in \cite{Kato}
but differs in that the poles appear in the tachyonic side in our case.) 
Then effectively there appears no singularity and the bound may
disappear, at the sacrifice of introduction of infinitely many 
tachyons. Using a Feynman rule with these propagators and vertices of
the BSFT action, it might be possible to reproduce correct string
amplitudes in a field-theoretical manner, also for the loop amplitudes.

There is another way to avoid this problem on the loops. In the previous
section, we found that the three-point function disappear once one of
the three momentum legs is put to be on-shell. If this is the general
feature of the $n$-point functions in the BSFT action evaluated with the
disk partition function, any field-theoretical loop amplitude
constructed from these vertices vanishes. Thus from the first place
there is no problem concerning the cut-off of the loop momenta. 
Then, how can we reproduce the string theory loop amplitudes from the
BSFT? A possible answer to this question might be that we have to
consider also the BSFT action based on partition functions of higher
genus worldsheets. (This standpoint is different from Witten's original 
proposal that the disk partition function is a definition of the BSFT.)
There are several problems even for one-loop BSFT's \cite{loopbsft},
which is beyond the study in this paper.  


\section{Conclusion and discussions}

The main virtue of the analytic continuation method which we have
developed in this paper is that it accommodates the BSFT and string
$\sigma$ models to the massive excitations. Allowing the normal-ordered
Fourier basis for the boundary couplings of the $\sigma$ model partition
function ($=$ the BSFT action), we can choose appropriate momentum
which manifestly makes the partition function finite. The analytic
continuation of the momenta brings us to any region of interest,
including especially the on-shell momenta.
The resulting BSFT off-shell two-point functions for the tachyon field,
the gauge field and some of the massive fields on a BPS/non-BPS D-brane
reproduce the well-known string mass-shell conditions. The BSFT
three-point function computed for two tachyons and a single gauge field 
on a brane-anti-brane provides the correct on-shell value of the 
standard scattering amplitude calculation. In doing this, 
we have used the field redefinition of the tachyon and the gauge 
field which makes the three-point function regular at the on-shell
value, instead of the prescriptions in \cite{Frolov, Coletti}. 
The three-point function obtained in section \ref{sec-3} is consistent
with the assumption $S_{\rm BSFT}=Z$ : since the partition function
already includes all the on-shell scattering amplitudes, the BSFT action
should not generate additional contributions to these from
one-particle-reducible Feynman graphs. In fact, our vertex vanishes if
we put one of the external legs to be on-shell. 

We have found that our BSFT tachyon action coincides with those
constructed so far in the Taylor (or derivative) expansion of
the tachyon boundary coupling or in the linear tachyon profiles. 
As for the two-point functions we need no field redefinition to relate
these, since the standard renormalization of the fields in the $\sigma$
model turns out to be identical with our renormalization based on the
normal-ordered Fourier basis. For the three-point function, a certain
field redefinition obtained by looking at the difference in gauge
transformations on both sides gives a perfect agreement of the on-shell
tachyon-tachyon-gauge interaction.

Because our analysis is based on the perturbation of 
the $\sigma$ model couplings for $T,A_{\mu}, \cdots$, 
how our methods may incorporate non-perturbative effects
of BSFT is a quite important subject to study. 
To illustrate this, let us recall the example \cite{LNT}
where a perturbative series in the BSFT sums up to give 
a neat result. For the rolling tachyon solution $T(X)=\lambda e^{-X_0}$  
the partition function was obtained perturbatively 
as $Z=\sum_n  (-(T(X_0))^2)^n/n$,
which can be analytically continued to $1/(1+T^2)$ to give the
information of the final state of the rolling tachyon \cite{LNT}.
This suggests that we need anyway the analytic continuation in general 
for the $\sigma$ model couplings. 
Then, how we can see directly non-perturbative effects of BSFT, like
instantons in gauge theories? An answer is found in
the evaluation of the partition function with the linear tachyon
profiles, or the constant gauge field strength, or more general cases
considered in \cite{li-witten}, which can be regarded as
non-perturbative results. 
The path integral of the world sheet action with these profiles
reduces to a Gaussian integral, and is evaluated exactly, 
thus the results are non-perturbative in the boundary couplings.
To find a more universal relation between this and the analytic
continuation method, and how to extend our analysis beyond the
perturbation, are interesting questions.

We would like to point out an intriguing similarity between our tachyon
two-point function (\ref{tacbsft}) and the partition function of the
linear tachyon profile (\ref{linearex}). In fact, ${\cal F}(x)$ in
(\ref{linearex}) is written as $2\sqrt{\pi} \Gamma(x+1)/\Gamma(x+1/2)$
which, as a function,  is the same as $Z^{(2)}(y)$ defined in 
(\ref{Z2tac2}) if we absorb the $2^{2y}$ factor into the redefinition of 
the tachyon field. In the former, the argument is 
$x = 2\alpha' (\p_\mu T)^2$ while in the latter $y$ is just
$\alpha'\p^2$. This surprising similarity suggests that there might be
some relation between these two, such as a certain field
redefinition. If this is true, our result could be applied to the
analysis of the rolling tachyon physics using BSFT \cite{ST, pre}. 

Although our approach have reproduced various perturbation
results of string theory, there are still many indispensable 
aspects which need to be explored to have a definite
``field-theoretical'' BSFT. 
First, we need to include space-time fermions. This is difficult because
we use the NS-R formalism, but it might be overcome by taking into
account the broken supersymmetry \cite{susy}. Secondly, general $n$-point
functions may have more complicated singularity structures, such as
cuts. We do not have tools powerful enough to compute explicitly the
higher-point functions which might exhibit interesting structures. At
least, the four-point functions should reproduce the Veneziano
amplitude and the $s$-$t$ channel duality. Lastly, the most important
is to understand the loop amplitudes of the BSFT discussed in the
previous section. These are very interesting questions and we hope to
come back to these in the future. 

\acknowledgments 

K.H.\ would like to thank M.~Kato for valuable discussions,
and Y.~Kikukawa and T.~Yoneya for comments. 
S.T.\ would like to thank J.~McGreevy, D.~Kutasov, A.~Parnachev,
S.~Sugimoto and T.~Takayanagi for useful discussions. 
The work of K.H.~was partly supported in part by the Grant-in-Aid for
Scientific Research (No.~12440060, 13135205, 15540256 and 15740143) from
the Japan Ministry of Education, Science and Culture.

\appendix

\section{Rolling tachyon solution}
\label{app:rolling}

In this appendix, we study the properties of a classical solution of the
tachyon equation of motion (\ref{taconshell}). Consider a solution of 
a plane wave with the momentum $k_0= i /\sqrt{2\alpha'}, k_i=0$, 
which is well promoted to be an exactly marginal operator \cite{stro,
LNT}, called a half S-brane. For this tachyon profile 
$T = \lambda e^{x^0/\sqrt{2\alpha'}}$, 
the energy and the pressure were computed in \cite{LNT} with an
arbitrary $\lambda$, by evaluating a disk amplitude with an insertion of
a single closed string vertex. Here we may compute those observables by
purely field-theoretical method in the sense of target space, since we
have an off-shell action for the tachyon. Let us see how our action
reproduce a part of the results of \cite{LNT}. We couple the system to
the gravity in a natural manner,  
\begin{eqnarray}
S = {\cal T}\int\!\! dx  \;
\sqrt{-g}\left[
1 -\frac12 
T(x)Z^{(2)}(-\overleftarrow{\nabla}_\mu \alpha'g^{\mu\nu}
\overrightarrow{\nabla}_\nu)
T(x)
+ {\cal O}(T^4)
\right] \ .
\end{eqnarray}
{}From this expression the energy is defined as 
 $T_{00} = [2\delta L / \delta g^{00}]_{g^{00}=-1}$. 
We may make a dimensional reduction to $1$ dimension without losing
generality. The reduced Lagrangian has an infinite number of derivatives
 (we define $ \sqrt{-g^{00}} \equiv  v$),  
\begin{eqnarray}
 L = {\cal T}\frac{1}{v}
\left[1 - \frac12
\sum_{n=0}^\infty a_n {\alpha'}^nv^{2n} (\nabla_0^n T)^2
\right] \ ,
\end{eqnarray}
thus it is nontrivial that the solution 
$T = \lambda e^{x^0/\sqrt{2\alpha'}}$
has a conserved energy, which we are going to check.  
The covariant derivatives are explicitly written as 
$\nabla_0^n T = (1/v^{n-1}) \p_0 (v\p_0(v\p_0(\cdots (v\p_0 T))))$.
Let us take a differentiation of $L$ with respect to $v$, giving 
the energy in the system, 
\begin{eqnarray}
\frac{1}{\cal T}\left. \frac{\delta L}{\delta v}\right|_{v=1}
&=& -\left(1 - \frac12 a_0 T^2\right) 
- \frac12 \sum_{n=1}^\infty a_n{\alpha'}^n
\frac{\delta}{\delta v} 
\left[\frac1{v}
\overbrace{(v\p_0 (v\p_0(v\p_0(\cdots (v\p_0}^{n}
T))))^2\right]_{v=1}\nonumber\\
&=&-\left(1 - \frac12 a_0 T^2\right)  - \frac12\sum_{n=1}^\infty
a_n{\alpha'}^n\sum_{i=1}^{2n-1} (\p_0^i T)(\p_0^{2n-i} T) (-1)^{i-n} \ . 
\label{befo}
\end{eqnarray}
The equation of motion $\sum_{n=0}^\infty a_n (-1/2)^n=0$ can be used
under the substitution of the solution 
$T = \lambda e^{x^0/\sqrt{2\alpha'}}$. 
We finally obtain the conserved energy 
\begin{eqnarray}
 T_{00} = {\cal T} \ , 
\end{eqnarray}
as we expected.

It is easy to show that the pressure $T_{ii}$, which can be computed in
the same manner from our Lagrangian, is identical to the 
partition function itself, since we are dealing with a spatially
homogeneous rolling. This pressure reproduces the small field expansion
of the result of the worldsheet computation \cite{LNT}, 
\begin{eqnarray} 
 T_{ii} = {\cal T}\left(
1 - \frac{\pi}2 
\left(\lambda e^{x^0/\sqrt{2\alpha'}}\right)^2  
+ {\cal O}(\lambda^4)\right) \ . 
\end{eqnarray}
We have used a relation $Z^{(2)}(y=1/2) = \sum_n a_n (1/2)^n = \pi$.
The pressure is decreasing, and the system is decaying to the tachyon
matter \cite{originalroll}. 

How about the the tachyon profile 
$T=\lambda \cosh ( {\bf X}^0/\sqrt{2\alpha'})$ which was originally
considered by Sen \cite{originalroll}? The computation of the partition
function faces a problem that $T(\tau)T(0)$ diverges as 
$\tau\rightarrow 0$ because of contributions from cross terms,
that is, $\langle :e^{X^0/\sqrt{2\alpha'}}(\tau)\!:
:e^{-X^0/\sqrt{2\alpha'}}(0)\!:\rangle$. Due to
this divergence, the integration over $\tau$ does not
converge. However, now we have a BSFT action using the analytic
continuation and it should be valid for this tachyon profile. In fact,
if we notice that the contributions from the cross terms are
proportional to the equation of motions and thus vanish, we can easily
check that the pressure and energy computed from the action agree with
those computed by Sen using the boundary state \cite{originalroll}
to the order we considered.

\section{Background constant field strength}
\label{app:constb}

Though the two-point functions obtained in this paper exhibit
interesting higher derivative structures, at least around the on-shell
momenta we can make a field redefinition to make them to be in a
canonical form. In this sense any nontrivial consequence on effective
action, except for the mass-shell conditions, may not come out from the
two-point functions. 
However, if we include nontrivial backgrounds and compute two-point
functions in those backgrounds, they contain information of higher point
functions in a reduced manner.  
In this appendix, we follow this line and adopt a constant gauge field
strength $\bar{F}_{\mu\nu}$ as a background. Many $\sigma$ model
calculations have been done so far on this background, which would be
good for comparison with our method. 

According to \cite{Andreev-Tseytlin}, the worldsheet boundary propagator
is  
\begin{eqnarray}
 \langle X^\mu(\varphi_1)X^\nu(\varphi_2)\rangle
&=& 4 \sum_{n=1}^{\infty} 
\frac1n e^{-\epsilon n} \left(
G^{\mu\nu} \cos(n\varphi_{12}) - i H^{\mu\nu} \sin (n\varphi_{12})
\right)
\nonumber 
\\
&  & \hspace{-20mm}= \; -2
\left[
(G^{\mu\nu}-H^{\mu\nu}) \log(1-e^{i\varphi_{12}-\epsilon})
+ 
(G^{\mu\nu}+H^{\mu\nu}) \log(1-e^{-i\varphi_{12}-\epsilon})
\right] \ , 
\nonumber
\\
\langle\psi^\mu(\varphi_1) \psi^\nu(\varphi_2)\rangle
&=& \frac{i}{2}
\left[
(H^{\mu\nu}-G^{\mu\nu})\frac{e^{(i\varphi_{12}-\epsilon)/2}}
{1-e^{i\varphi_{12}-\epsilon}}
+(H^{\mu\nu}+G^{\mu\nu})\frac{e^{(-i\varphi_{12}-\epsilon)/2}}
{1-e^{-i\varphi_{12}-\epsilon}}
\right]
\end{eqnarray}
where
\begin{eqnarray}
 G^{\mu\nu} = (1-\bar{F}^2)^{-1} \ , \quad 
  H^{\mu\nu} = \bar{F}(1-\bar{F}^2)^{-1} \ .
\end{eqnarray}
Here $\bar{F}$ is a background constant field strength, and $G_{\mu\nu}$
is so-called open string metric.
Let us first consider a self-contraction to give a renormalized 
boundary coupling. 
The boundary coupling is the same as before, (\ref{gaugeintc}).
The redefinition of the gauge field to absorb the divergent factor
coming from the normal ordering is
\begin{eqnarray}
 a_\mu (k) 
= a_{{\rm R}\mu(k)}
\exp[2k_\mu G^{\mu\nu} k_\nu \log \epsilon] \ .
\end{eqnarray}
As in section \ref{subsec-two-point gauge}, expansion of this and
addition of a total derivative term gives
\begin{eqnarray}
 a_\rho (k) &=& a_{{\rm R}\rho(k)} + 
2k_\mu G^{\mu\nu} k_\nu (\log \epsilon)a_{{\rm R}\rho(k) }
- 2k_\mu G^{\mu\nu} k_\rho (\log \epsilon)a_{{\rm R}\nu(k) }
\nonumber \\
&=& a_{{\rm R}\rho(k)} - 
2ik_\mu G^{\mu\nu} (\log \epsilon)f_{{\rm R}\nu\rho}(k) \ .
\end{eqnarray}
This renormalization is the same as that of \cite{Andreev-Tseytlin}.

A crucial difference from the boundary coupling in section
\ref{subsec-two-point gauge} appears in the self-contractions. 
The term (\ref{self-cont}), which vanished for the trivial background,
is now giving a nonzero contribution: 
\begin{eqnarray}
\int\! d\tau  
\langle \dot{X}^\mu(\tau) X^\sigma(\tau)\rangle 
ik^\sigma : e^{ik\cdot \hat{X}} :
= \frac{1}{e^\epsilon-1} 4H^{\mu\sigma} ik_\sigma
\int\! d\tau : e^{ik\cdot \hat{X}} : \ .
\end{eqnarray}
In addition, the fermion self-contraction is also non-vanishing as 
\begin{eqnarray}
\langle\psi^\mu(\varphi) \psi^\nu(\varphi)\rangle
&=& i H^{\mu\nu} \frac{e^{\epsilon/2}}{e^\epsilon-1} \ .
\end{eqnarray}
Thus the diverging part of the self-contraction cancels with each other,
while a finite term remains :
\begin{eqnarray}
 I_{\rm B} &=& -i\!\!
\int\!\! d\tau\!\! \int\!\! dk \left(
a_{{\rm R}\mu} (k) :\dot{X}^\mu e^{ik\cdot X}\!\!:
-2f_{{\rm R}\mu\nu}(k) :\psi^\mu \psi^\nu e^{ik\cdot X}\!:
-iH^{\mu\nu}f_{{\rm R}\mu\nu}(k) :e^{ik\cdot X}\!:
\right) \ .
\nonumber
\end{eqnarray}
The last term is the finite modification due to the renormalization
in the presence of the background constant field strength.

Interestingly, this modification results in non-vanishing one-point
function, 
\begin{eqnarray}
 \left\langle I_{\rm B}\right\rangle = - H^{\mu\nu} F_{{\rm R}\mu\nu}(x) 
\ .
\end{eqnarray}
This is expected, as an expansion of the Maxwell Lagrangian around
a constant field strength.\footnote{This existence of the one-point
function does not cause any problem as opposed to the situation in
section \ref{subsub24}, because it is proportional to the field strength
and thus it does not change the vacuum and the constant field strength  
is a solution of the equations of motion. } So our renormalized boundary
coupling is consistent with the usual target space picture.

Let us compute the two-point function with this boundary coupling. 
A straightforward calculation shows that 
\begin{eqnarray}
\langle  I_{\rm B}I_{\rm B}\rangle\!
= \int\!\! dkd\widetilde{k}\; e^{i(k + \widetilde{k})\cdot x}
8\pi^{3/2}2^{4k_\mu G^{\mu\nu}\widetilde{k}_\nu}
\frac{\Gamma(2k_\mu G^{\mu\nu} \widetilde{k}_\nu+1/2)}
{\Gamma(2k_\mu G^{\mu\nu} \widetilde{k}_\nu+1)}
f_{\rho\sigma}(k) f_{\delta\gamma}(\widetilde{k}) 
G^{\rho\delta} G^{\sigma\gamma} \ .\;\;\;\;\;\;
\label{ourde}
\end{eqnarray}
The higher-derivative part turns out eventually to be 
nothing different from the one obtained in
the case of vanishing field strength background, except that the metric
is now replaced by the open string metric $G_{\mu\nu}$. 
Our result (\ref{ourde}) is consistent with the partition function
results of \cite{Andreev-Tseytlin}. However, our result is slightly
different from the effective action derived from string
scattering amplitudes provided in \cite{Andreev-Tseytlin}. 
This discrepancy might be resolved by some field redefinitions or
Jacobi-like identities among field strengths.


\newcommand{\J}[4]{{\sl #1} {\bf #2} (#3) #4}
\newcommand{\andJ}[3]{{\bf #1} (#2) #3}
\newcommand{\AP}{Ann.\ Phys.\ (N.Y.)}
\newcommand{\MPL}{Mod.\ Phys.\ Lett.}
\newcommand{\NP}{Nucl.\ Phys.}
\newcommand{\PL}{Phys.\ Lett.}
\newcommand{\PR}{ Phys.\ Rev.}
\newcommand{\PRL}{Phys.\ Rev.\ Lett.}
\newcommand{\PTP}{Prog.\ Theor.\ Phys.}
\newcommand{\hep}[1]{{\tt hep-th/{#1}}}

\end{document}